\documentclass{emulateapj}
\usepackage{natbib}
\usepackage{graphicx}
\usepackage{url}
\usepackage{color}
\usepackage{gensymb}

\begin{document}

\title{Photometric Redshifts with the LSST II: \\ The Impact of Near-Infrared and Near-Ultraviolet Photometry}

\author{Melissa~L.~Graham\altaffilmark{1}}
\author{Andrew~J.~Connolly\altaffilmark{1}}
\author{Winnie~Wang\altaffilmark{1}}
\author{Samuel~J.~Schmidt\altaffilmark{2}}
\author{Christopher~B.~Morrison\altaffilmark{1}}
\author{\v{Z}eljko~Ivezi\'{c}\altaffilmark{1}}
\author{S\'{e}bastien~Fabbro\altaffilmark{3}}
\author{Patrick~C\^{o}t\'{e}\altaffilmark{3}}
\author{Scott~F.~Daniel\altaffilmark{1}}
\author{R.~Lynne~Jones\altaffilmark{1}}
\author{Mario~Juri\'{c}\altaffilmark{1}}
\author{Peter~Yoachim\altaffilmark{1}}
\author{J.~Bryce~Kalmbach\altaffilmark{1}}
\altaffiltext{1}{DiRAC Institute, Department of Astronomy, University of Washington, Box 351580, U.W., Seattle WA 98195}
\altaffiltext{2}{Department of Physics, UC Davis, One Shields Avenue, Davis CA 95616}
\altaffiltext{3}{National Research Council of Canada, Herzberg Astronomy \& Astrophysics Research Centre, 5071 W. Saanich Rd, Victoria, BC V9E 2E7}

\begin{abstract}
Accurate photometric redshift (photo-$z$) estimates are essential to the cosmological science goals of the Vera C. Rubin Observatory Legacy Survey of Space and Time (LSST). In this work we use simulated photometry for mock galaxy catalogs to explore how LSST photo-$z$ estimates can be improved by the addition of near-infrared (NIR) and/or ultraviolet (UV) photometry from the Euclid, WFIRST, and/or CASTOR space telescopes. Generally, we find that deeper optical photometry can reduce the standard deviation of the photo-$z$ estimates more than adding NIR or UV filters, but that additional filters are the only way to significantly lower the fraction of galaxies with catastrophically under- or over-estimated photo-$z$. For Euclid, we find that the addition of {\it JH} $5{\sigma}$ photometric detections can reduce the standard deviation for galaxies with $z>1$ ($z>0.3$) by ${\sim}20\%$ (${\sim}10\%$), and the fraction of outliers by ${\sim}40\%$ (${\sim}25\%$). For WFIRST, we show how the addition of deep {\it YJHK} photometry could reduce the standard deviation by ${\gtrsim}50\%$ at $z>1.5$ and drastically reduce the fraction of outliers to just ${\sim}2\%$ overall. For CASTOR, we find that the addition of its ${UV}$ and $u$-band photometry could reduce the standard deviation by ${\sim}30\%$ and the fraction of outliers by ${\sim}50\%$ for galaxies with $z<0.5$. We also evaluate the photo-$z$ results within sky areas that overlap with both the NIR and UV surveys, and when spectroscopic training sets built from the surveys' small-area deep fields are used. 
\end{abstract}

\section{Introduction} \label{sec:intro}

Photometric redshifts (photo-$z$'s; for a broad review see, e.g., \citealt{2019NatAs...3..212S}) are a key component to achieve the cosmological science goals of the Vera C. Rubin Observatory Legacy Survey of Space and Time (LSST), including weak lensing, galaxy clusters, and supernova host galaxies \citep[e.g.,][]{2018RPPh...81f6901Z,2019A&A...623A..76A,Ivezi__2019}. Establishing how to obtain the most accurate photo-$z$ possible from the LSST data set is currently an active and urgent area of study. Towards this goal, \citet[][paper I in this series]{2018AJ....155....1G} presented and characterized the color-matched nearest-neighbors (CMNN) photometric redshift estimator and demonstrated its use as an efficient tool to evaluate how potential changes to LSST survey parameters directly impact the LSST photo-$z$ results.  \citet{2020arXiv200103621S} provides a comparison of the CMNN technique to eleven other modern photo-$z$ algorithms using a variety of metrics and mock LSST galaxy catalogs. 

Increasing the wavelength coverage of a galaxy catalog's photometric data can improve photometric redshift estimates \citep[e.g.,][]{2008MNRAS.386.1219B,2010A&A...523A..31H}. This is because doing so provides additional information about the galaxy's spectral energy distribution (SED). For example, when only optical filters are available, galaxies at redshifts of $z\sim0.1$ have similar colors to galaxies at $z\sim2$. This is because the change in color due to the passage of the Balmer break (at $4000$ $\rm \AA$) through the $u$ band filter as a function of redshift is nearly indistinguishable from the change in color due to the Lyman alpha break (at $912$ $\rm \AA$). This degeneracy causes catastrophic outliers, as photo-$z$ estimators have trouble distinguishing between very low and very high-$z$ galaxies. This issue is mitigated if we can jointly track the Balmer and Lyman breaks as their observed wavelength moves with redshift by extending the filter coverage into the UV and NIR. 

\citet{2018AJ....155....1G} used the CMNN estimator to illustrate how the LSST photo-$z$ quality would deteriorate in regions without $u$- or $y$-band coverage; for example, the standard deviation in $z_{\rm true}-z_{\rm phot}/(1+z_{\rm phot})$ increases by ${\sim}$50\% at intermediate redshifts ($z_{\rm phot}\approx1.5$). In this work, we use the CMNN estimator to evaluate the benefit of adding near-infrared (NIR) and ultraviolet (UV) photometry to the LSST optical data when estimating galaxy photometric redshifts. We consider two potential sources of NIR photometry: the European Space Agency (ESA) Euclid mission and the National Aeronautics and Space Administration (NASA) Wide-Field InfrarRed Survey Telescope (WFIRST). We also consider one potential source of UV photometry, a proposed satellite led by the Canadian Space Agency (CSA): the Cosmological Advanced Survey Telescope for Optical and ultraviolet Research (CASTOR\footnote{``Castor" is the genus of, and french word for, ``beaver". The North American beaver, {\it Castor canadensis}, is the national animal of Canada.}).

ESA's Euclid mission \citep{2011arXiv1110.3193L,2016ASPC..507..401J,2016SPIE.9904E..0OR} is a planned $7$ year NIR survey, which (at the time of this analysis) is set to launch in 2021, ${\sim}2$ years before the start of LSST operations. Euclid will cover $15000$ square degrees of sky: nearly the entire extragalactic sky with Galactic latitude $>|30|$, and an avoidance zone around the ecliptic as well. Its step-and-stare survey strategy will cover the full area to a depth of {\it YJH} $\sim24$ magnitudes (AB magnitudes, for a $5\sigma$ point source; \citealt{2016ASPC..507..401J}). Revisits to one or more ${\sim}40$ square degree ``deep field'' will extend this limit by ${\sim}2$ magnitudes. By 2028, approximately 7000 square degrees (${\sim}40\%$) of the LSST's ``wide-fast-deep" (WFD) main survey area will have overlapping NIR coverage from Euclid (see, e.g., Figure 6 of \citealt{2017ApJS..233...21R}). The addition of Euclid's $J$ and $H$ filters are expected to improve photo-$z$ estimates at and above redshift ${\sim}1.5$, where the $4000$ $\rm \AA$ Balmer break is redshifted to $10000$ $\rm \AA$ and begins to influence the $Y-J$ color. 

NASA's WFIRST \citep{2015arXiv150303757S} will produce a sky survey that has ${\sim}2200$ square degrees of overlap ($3.8\times10^8$ galaxies) with the LSST WFD main survey and $5\sigma$ point-source detection limits down to {\it YJH} $\gtrsim26.5$ magnitude. WFIRST will also survey in a redder band (F184, $1.7-2.0$ $\rm \mu m$; referred to as $K$ for simplicity in this work) to a shallower depth ($\gtrsim25.5$). WFIRST is currently planned for a mid-2020s launch date. The addition of WFIRST's {\it YJH} filters are expected to improve photo-$z$ estimates for the same reason as Euclid, sampling the Balmer break at intermediate redshifts. Additionally, the redder filter at $1.7-2.0$ $\rm \mu m$ should improve photo-$z$ estimates above redshift ${\sim}2.5$, where the Balmer break influences the $H-K$ color.

The scientific benefits of combining data from LSST, Euclid, and WFIRST has been amply demonstrated in the literature. For example, \citet{2017ApJS..233...21R} investigate a variety of science goals that benefit from the combination of LSST and Euclid data at either the pixel or catalog level. They show that improvements to the photo-$z$ accuracy will improve the weak lensing signal in particular, but also be of benefit to classifying high-$z$ supernovae and making cluster mass estimates. As another example, \citet{2015arXiv150107897J} shows that when WFIRST NIR photometry is included in LSST photo-$z$ estimates, the most significant improvements are reductions in the scatter between photometric ($z_{\rm phot}$) and true ($z_{\rm true}$) redshifts for galaxies with $z_{\rm phot}>1.5$, and in the catastrophic outliers rate across the full range of $0.0<z_{\rm phot}<3.5$. They discuss how this improvement is likely to have an impact on a wide variety of extragalactic and cosmological science goals.

The proposed CASTOR mission \citep{2012SPIE.8442E..15C,2019BAAS...51g.219C}, led by the CSA with international partners, aims to carry out wide-field imaging in three passbands located in the UV/blue-optical region (0.15--0.55 $\mu$m). The depths and areal coverage of these surveys are currently under investigation, but for the purposes of our analysis we consider two notional surveys: (1) the Primary Survey, a ${\sim}7200$ square degree field in the north that overlaps with the Euclid-Wide survey (but possibly has a limited overlap with LSST); and (2) the Cadence Survey, ${\sim}20$ square degrees in the south that overlaps with WFIRST and LSST, in which depth is built up over many repeated visits. The $5{\sigma}$ limiting magnitudes for CASTOR Primary would be $m_{\rm UV} \sim 27.4$, $m_u \sim 27.4$, and $m_g \sim 27.1$ mag, and for CASTOR Cadence they would be $m_{\rm UV} \sim 29.25$, $m_u \sim 28.95$, and $m_g \sim 28.45$ mag \citep{2014SPIE.9144E..03C}. There are currently two proposed CASTOR passband pairs, $UV$ and $u$, and $UV$-dark and $u$-wide; we consider both pairs in our analysis. The addition of CASTOR $UV$ and deep $u$-band photometry is expected to improve the photo-$z$ estimates for low-redshift galaxies ($z<1.5$) by breaking the color degeneracy between low- and high-redshift galaxies, and also by sampling the UV-upturn from the blackbody emission of evolved hot stars at $\gtrsim1000$ $\rm \AA$ \citep[e.g.,][]{1990ApJ...364...35G,1999MmSAI..70..691G}, and the Lyman-$\alpha$ break at $912$ $\rm \AA$.

In this work, we focus on how the combination of LSST, Euclid, WFIRST, and/or CASTOR photometry at the catalog level will impact photo-$z$ estimates at low ($0.3 < z \lesssim 0.6$), intermediate ($z \sim 1$), and high ($z>2$) redshifts. The LSST Science Requirements Document \citep[SRD;][]{LPM-17} defines the minimum deliverables for statistical measures of the accuracy of photometric redshifts derived from LSST photometry. First, the root-mean-square error in photo-z must be $< 0.02(1 + z_{\rm phot})$; second, the $3{\sigma}$ outlier fraction must be $< 10\%$; and third, the average bias must be $< 0.003(1 + z_{\rm phot})$. These target values apply to an $i<25$ mag sample of $4\times10^9$ galaxies from $0.3 < z < 3.0$, and serve as suitable reference targets for this work. We therefore limit our analysis to a simulated set of galaxies with $i<25$ mag and $0.3 < z < 3.0$, and include these target values as a reference benchmark in our analyses\footnote{Some readers might notice that the definition of a LSST ``Gold Sample" has evolved from $i<25.3$ mag \citep{LPM-17}.}.

In Section \ref{sec:exp} we describe our simulation design: the simulated galaxy catalog, its optical, NIR, and UV photometry, our photo-$z$ estimator, and how we analyze the results. In Sections \ref{sec:euclid} and \ref{sec:wfirst} we evaluate the impact of including NIR photometry from Euclid and WFIRST, respectively, and in Section \ref{sec:castor} we demonstrate the impact of including UV, $u$- and $g$-band photometry from CASTOR. We summarize and discuss our results in Section \ref{sec:conc}, and present several ideas for additional ways in which the space-based imaging of these missions can provide further improvements to the LSST photo-$z$ estimates.

\section{Simulation Design}\label{sec:exp}

In order to evaluate the impact of adding NIR and/or UV to the LSST optical photometry to estimated photometric redshifts, we use the color-matched nearest-neighbors (CMNN) photo-$z$ estimator (Section \ref{ssec:exp_cmnn}) with a simulated galaxy catalog with realistic observed photometric characteristics (Section \ref{ssec:exp_cat}). Section \ref{ssec:exp_stat} describes the statistical measures and evaluation tools that we will use in our analysis.

\subsection{The CMNN Photo-$z$ Estimator}\label{ssec:exp_cmnn}

The CMNN photometric redshift estimator that we use for this work was first introduced and described by \citet{2018AJ....155....1G}\footnote{A demonstrative version of the CMNN estimator is available at \url{https://github.com/dirac-institute/CMNN\_Photoz\_Estimator}.}. It is not intended to provide the ``official'' or ``best'' LSST photo-$z$, but instead to provide sets of photo-$z$ results in which the accuracy and precision of the photo-$z$ estimates are directly related to the precision of the input photometry. In this way, it is particularly useful for comparing the {\em relative} photo-$z$ quality from different survey strategies that affect the LSST photometric quality of the LSST.

The CMNN photo-$z$ estimator requires ``training" and ``test" galaxy catalogs: the former is equivalent to a set of galaxies with ``known" or spectroscopic redshifts, and the latter to a sample of galaxies for which photo-$z$ will be estimated. The full catalog of simulated galaxies that we use for this work is described in Section \ref{ssec:exp_cat}. For this work, instead of splitting the simulated catalog into test and training sets we employ a leave-one-out strategy: every catalog galaxy is considered as the test galaxy in turn, and the training set is composed of all the other galaxies in the simulated catalog (Section \ref{ssec:exp_cat}). For the test galaxy, the estimator identifies a subset of training galaxies that are its nearest neighbors in color-space using the Mahalanobis distance\footnote{This is an approximation of the Mahalanobis Distance, which is typically expressed as $D_M^2 = (\bar{x}-\bar{u})^T S^{-1} (\bar{x}-\bar{u})$ where $S$ is the covariance error matrix. For this work we verified that using the approximation does not significantly change the photo-$z$ estimates, and has no impact on our overall conclusions.}: 

\begin{equation}\label{eq:DM}
D_M = \sum_{\rm 1}^{N_{\rm colors}} \frac{( c_{\rm train} - c_{\rm test} )^2}{ (\delta c_{\rm test})^2},
\end{equation}

\noindent where $c_{\rm train}$ is the color of a training set galaxy, $c_{\rm test}$ is the color of a test set galaxy, and $\delta c_{\rm test}$ is the measurement error in the color of the test set galaxy (Section \ref{sssec:exp_cat_err}). If a galaxy is undetected in a given filter, the associated color does not contribute to $D_M$. We enforce that the number of colors be $N_{colors} \geq 3$ (i.e., a galaxy must be detected with at least 3 colors to obtain a photo-$z$ estimate). 

Once $D_M$ is measured for all training-set galaxies, we use a percent-point function value of $PPF=0.68$ and the degrees of freedom, $N_{colors}$, to define a threshold value, $\tau_{D_M}$. Training-set galaxies with $D_M$ values below $\tau_{D_M}$ are identified as the set of {\it color-matched nearest-neighbors} (CMNN). For example, for $N_{\rm colors}=5$, PPF$=68$ per cent of all training galaxies consistent with the test galaxy will have $D_M < 5.86$. The CMNN subset galaxies are then subjected to a weighted ($D_M^{-1}$)\footnote{Since the test galaxy is left out of its own training set, there are no instances of $D_M=0$ and no infinite weights.} random selection, and the redshift of the chosen training-set galaxy is adopted as the photo-$z$ for the test galaxy. For test galaxies with less than $10$ training-set galaxies in their CMNN subset ($N_{CM}<10$), the $10$ nearest neighbors are used. If $N_{\rm CM}\geq10$, the photo-$z$ uncertainty ($\delta z$) is the standard deviation in the true redshifts of the color-matched subset; if $N_{\rm CM}<10$, we calculate what PPF threshold value associated with the $10^{th}$ nearest-neighbor, and inflate the uncertainty by $\times {\rm PPF_{10}}/0.68$.

Note that this is different from the application in \citealt{2018AJ....155....1G}, which also used a $PPF=0.68$ but chose the nearest neighbor in color-space, and did not have a modification for galaxies with $N_{CM}<10$ matches (such that galaxies with zero matches simply failed to obtain a photo-$z$). We furthermore do not apply an optical magnitude or color ``pseudo-prior" to our photo-$z$ estimates, as done in \citealt{2018AJ....155....1G}, so as to not give the optical an additional influence on the photo-$z$ results in this paper, which focuses on the addition of UV and NIR photometry. The other major difference compared to \citealt{2018AJ....155....1G} is that here, the entire simulated set of galaxies is used as both test and training set instead of assigning each galaxy to one or the other before simulating photo-$z$ results. This has allowed us to use much larger test sets and achieve more accurate statistical measures for the photo-$z$ results in high-redshift bins.

Incorporating the NIR filter(s) and/or UV passbands causes an increase in $N_{\rm colors}$, which increases the number of degrees of freedom in Equation \ref{eq:DM}. This can also increase the size of the CMNN subset of training galaxies and, at times, degrade the overall photo-$z$ quality. This is an important aspect of the CMNN estimator to keep in mind as we explore the impact of adding NIR and UV photometry. The photometric quality of the data must be good enough to overcome the increase in degrees of freedom in the estimator in order to see a beneficial impact on the photo-$z$ results. In this work we encounter several instances where the photometric quality for some simulated galaxy populations is insufficient to benefit from the addition of NIR or UV filters, and the photo-$z$ results are deteriorated. Generally we will mitigate this issue by identifying these populations using their observed characteristics (i.e., not using true catalog redshift), and then excluding the additional filters from the photo-$z$ estimates. Each of these instances is discussed in detail in the relevant analysis section.

\subsection{Simulated Galaxy Catalog}\label{ssec:exp_cat}

As in \citet{2018AJ....155....1G}, we use a galaxy catalog based on the Millennium simulation \citep{2005Natur.435..629S} -- specifically, a catalog based on the galaxy formation models of \citet{2014MNRAS.439..264G} and fabricated using the lightcone construction techniques described by \citet{2013MNRAS.429..556M}\footnote{Documentation for this catalog can be found at \url{http://galaxy-catalogue.dur.ac.uk}}. This simulated galaxy catalog was designed to model the optical and NIR properties of galaxies, and with appropriate limits also serves as a realistic representation of future LSST catalogs. ``True" catalog apparent magnitudes for optical and NIR filters {\it ugrizy} and {\it YJHK} are included in this simulated galaxy catalog (for simplicity, we refer to WFIRST's F184 filter as $K$ in this work because it fulfills a similar role), but we must synthesize magnitudes to represent the proposed CASTOR passbands.

\subsubsection{CASTOR Passbands and Simulated Photometry}\label{sssec:exp_cat_castor}

\begin{figure}
\begin{center}
\includegraphics[width=8.5cm,trim={0cm 0cm 0cm 0cm}, clip]{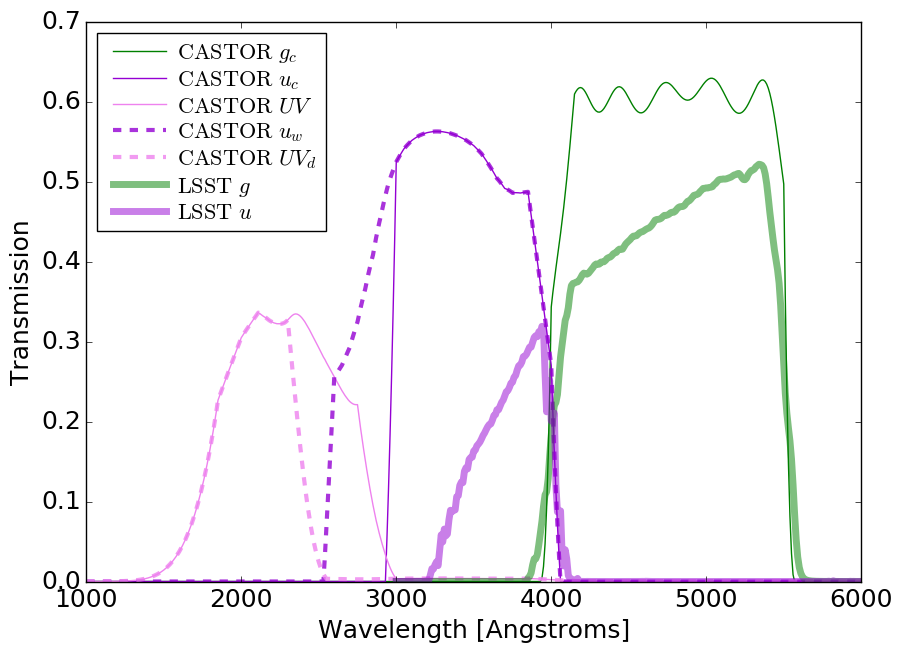}
\includegraphics[width=8.5cm,trim={0cm 0cm 0cm 0cm}, clip]{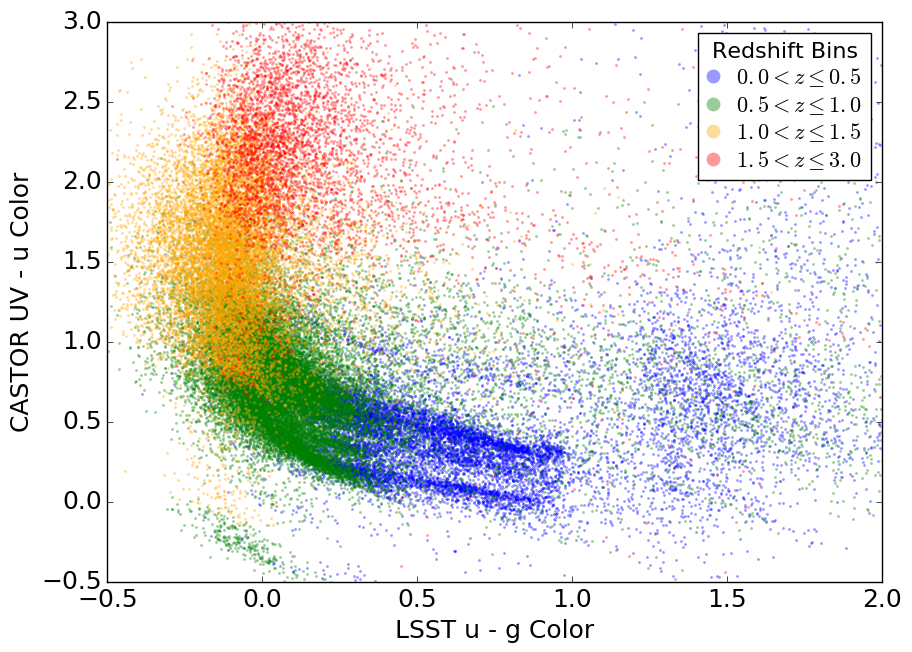}
\caption{{\it Top:} The full system transmission functions for the CASTOR $UV$, $UV_d$, $u_c$, $u_w$, and $g_c$ bandpasses considered in this work, compared to the LSST $u$ and $g$ filters. {\it Bottom:} Observed photometric colors for our simulated galaxies: CASTOR $UV$ $-$ LSST $u$ {\it vs.} LSST $u-g$. Points are colored by true catalog redshift (as in legend). Galaxies are clearly grouped by redshift in this 2d plane of UV-color {\it vs.} optical-color, which confirms to us that the UV photometry we have synthesized will provide some redshift information to the CMNN photo-$z$ estimates. A random $50\%$ of catalog galaxies are included in this plot, for clarity; the banding is due to the finite number of SED templates. \label{fig:nuv}}
\end{center}
\end{figure}

To simulate ``true" catalog photometry for the CASTOR passbands, we start by using the same process as in \citet[][their Section 5.1]{2018AJ....155....1G} to assign each galaxy a SED type from the catalog of \citet{2014ApJS..212...18B}. First, we find the SED template that best matches each catalog galaxy's rest-frame $g-r$ color. The SEDs are sequenced from early- to late-type, numbered from $0$ to $128$, and we include a modest amount of intrinsic scatter by simply adding a small uniform-random integer to the assigned SED number (within $\pm 2$). Next we flux-scale the SED to match the catalog's LSST $u$- and $g$-band apparent magnitudes, and then apply the four CASTOR bandpasses\footnote{The four potential CASTOR transmission functions that we use are the latest and most realistic estimates as of December 2019.} (see Figure \ref{fig:nuv}) to calculate ``true" catalog magnitudes in $u_c$, $u$-wide ($u_w$), $UV$, and $UV$-dark ($UV_d$). We do not recalculate a CASTOR $g_c$-band true catalog magnitude because the passband shape is quite similar to the LSST $g$ filter, as shown in the top panel of Figure \ref{fig:nuv}. Instead, we will simply apply the CASTOR magnitude limits to the catalog's LSST $g$-band apparent magnitudes.

To validate that our synthesized CASTOR $UV$ magnitudes are providing independent information about redshift, we plot the simulated galaxies in the color-color plane of CASTOR $UV$ $-$ LSST $u$ {\it vs.} the LSST $u-g$ in the bottom panel of Figure \ref{fig:nuv}. Plotted point color represents the true catalog redshift for each galaxy, and we can see that redshift is correlated with the $UV$-$u$ color in a way that is unique from the $u$-$g$ color. This confirms that our synthesized CASTOR UV photometry should have an influence on the estimated photometric redshifts. Since the $UV$ magnitudes are the only passband being synthesized in this work, we do not show similar color-color plots for Euclid or WFIRST NIR. The impact of CASTOR photometry on LSST photo-$z$ is presented in Section \ref{sec:castor}.

\subsubsection{Photometric Errors}\label{sssec:exp_cat_err}

\begin{figure}
\begin{center}
\includegraphics[width=8.5cm,trim={0cm 0cm 0cm 0cm}, clip]{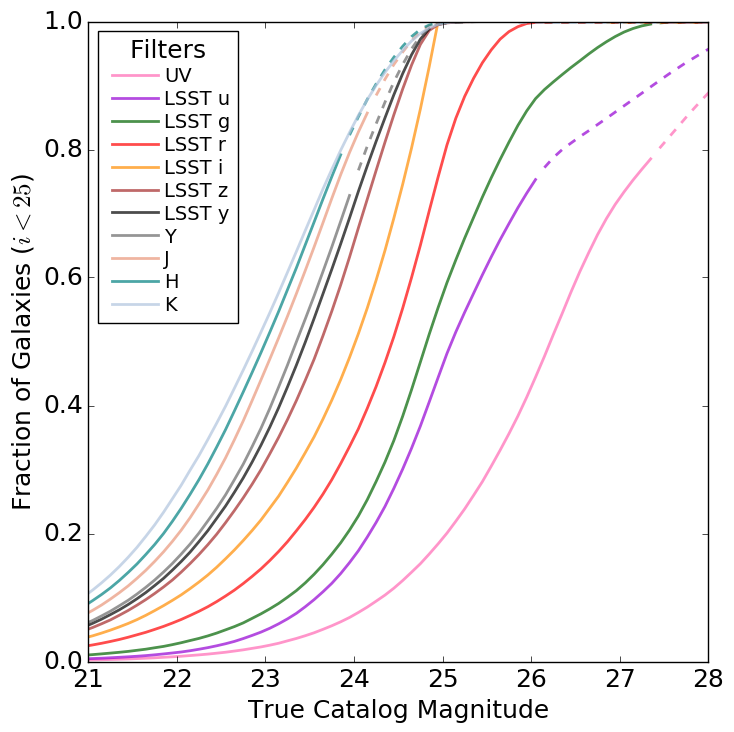}
\includegraphics[width=8.5cm,trim={0cm 0cm 0cm 0cm}, clip]{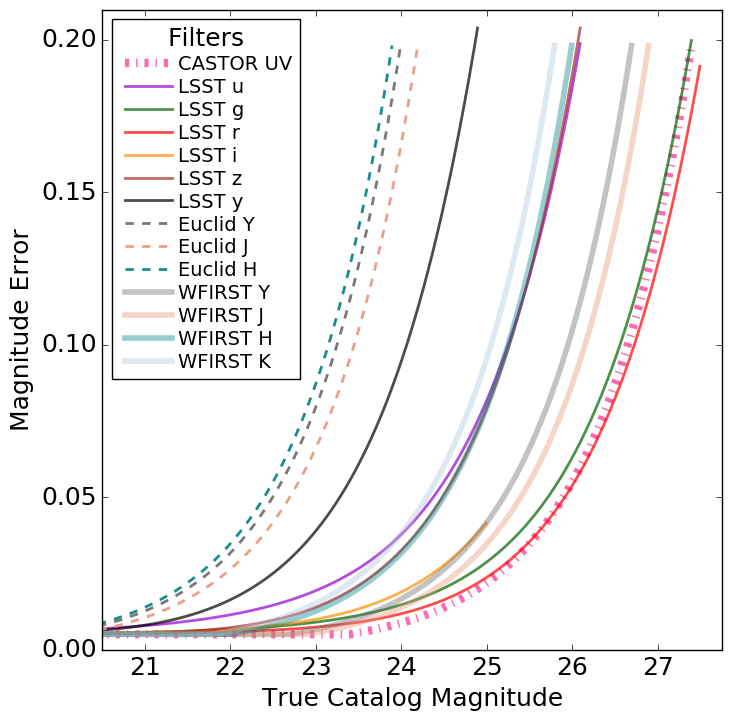}
\caption{{\it Top:} The cumulative distribution of true catalog apparent magnitudes for the set of simulated galaxies, with the imposed constraints of $i<25$ and $0.3<z<3.0$, for each of the six LSST filters, {\it ugrizy}, the four NIR filters considered in this work, {\it YJHK}, and the CASTOR UV passband. The dashed portions of the lines mark the range of magnitudes that are below the LSST, Euclid, and CASTOR surveys' $5\sigma$ detection limits (Table \ref{tab:maglims}). Note that this set of galaxies would be completely detected by WFIRST. {\it Bottom:} Predicted magnitude error {\it vs.} the catalog apparent magnitude for the six LSST bands {\it ugrizy}, the Euclid photometry of NIR bands {\it YJH} (dashed), the WFIRST photometry for NIR bands {\it YJHK}, and the CASTOR $UV$ band, each to the surveys' $5\sigma$ limiting magnitudes. \label{fig:cats_m}}
\end{center}
\end{figure}

To incorporate an LSST-like observational uncertainty into our catalog we simulate observed apparent magnitudes from the true catalog magnitudes by adding a normal random scatter\footnote{The normal random selection of the amount of scatter to apply, which is based on the simulated photometric error, is done in magnitude-space for all galaxies.} with a standard deviation equal to the predicted magnitude error for each galaxy. Predicted magnitude errors for the LSST, as described in Section 3.2.1 of \citet{Ivezi__2019}, depend on the galaxy's magnitude and the total survey exposure time elapsed in a given filter (for this work, we assume no additional components from e.g., deblending, or the different image resolutions of the four facilities). We assume standard observing conditions, an airmass of $1.2$, and a uniform survey progression that accumulates a total of 56, 80, 184, 184, 160, and 160 visits in filters {\it ugrizy} by year 10 (where each visit is 30 seconds of integration time). We use the LSST simulations software package described by \citet{2014SPIE.9150E..14C} to calculate the magnitude errors; it includes a systematic floor of $0.005$ in all filters. After 10 years of the LSST survey, these accumulated visits result in $5\sigma$ detection limits for each filter of $u<26.1$, $g<27.4$, $r<27.5$, $i<26.8$, $z<26.1$, and $y<24.9$ (as also listed in Table \ref{tab:maglims}). In the bottom panel of Figure \ref{fig:cats_m} we show the magnitude error as a function of observed apparent magnitude, with magnitude cuts applied at the $5{\sigma}$ limiting magnitude in filters {\it ugrzy} and at $i\leq25$ mag.

To predict the photometric errors for the near-infrared filters of Euclid and WFIRST, and the UV/blue passbands of CASTOR, we use the expected $5\sigma$ limiting magnitudes listed in Table \ref{tab:maglims}. We note that the predicted WFIRST limiting magnitudes have slightly evolved since our simulations were run. \citet{2019BAAS...51c.341D} reports that the expected $5{\sigma}$ point source imaging depths in $Y$, $J$, $H$, and $K$ (F184) are $26.9$, $26.95$, $26.9$, and $26.25$, respectively -- not too far off from the values we have used (except for $H$-band). For all three future facilities, we use the same prescription for error as described in Section 3.2.1 of \citet{Ivezi__2019}: 

\begin{equation}\label{eq:merr}
\sigma^2_{\rm rand} = (0.04-\gamma)x + \gamma x^2, 
\end{equation}

\noindent where $x=10^{0.4(m-m_5)}$, $m_5$ is the $5\sigma$ limiting magnitude, $m$ is the magnitude of the galaxy, and we use the fiducial value of $\gamma=0.04$ which sets the impact of, e.g., sky brightness, to be zero. As with the LSST photometry, we impose a systematic floor of $\sigma_{\rm rand} \geq 0.005$ mag. The magnitude error as a function of apparent magnitude, down to the $5\sigma$ detection limit of each filter for the surveys considered in this work, is shown in the bottom panel of Figure \ref{fig:cats_m}. 

The uncertainty in the apparent observed color of a galaxy is calculated as the root of the sum of the squares of the magnitude uncertainties in the two filters (i.e., the magnitude uncertainties from the two filters are added in quadrature, under the assumption of uncorrelated errors). In some parts of our analysis we extend the cut on apparent magnitude from the $5$ to the $1\sigma$ limiting magnitude, and in all cases we use $m_{\rm 1\sigma} = m_{\rm 5\sigma} + 1.75$ mag. In practice, obtaining matched aperture photometry for images with very different resolutions can lead to additional uncertainties in the combined catalogs' photometry (e.g., see Section 5.1 in \citealt{2015arXiv150303757S}). For this work we must assume that this potential issue has been resolved by a sophisticated joint pixel analysis.

\begin{table}
\begin{center}
\caption{$SNR=5$ Limiting Magnitudes}
\label{tab:maglims}
\begin{tabular}{lcccccccccc}
\hline
\hline
Filter     & LSST\footnote{Limits for a nominal 10-year survey.} & Euclid\footnote{Limits from Jean-Charles Cuillandre, private conversation.} & WFIRST\footnote{Limits from B. Jain, private conversation.} & CASTOR\footnote{Limits for the Primary (wide-area) survey.}\\
$UV$    & $\ldots$ & $\ldots$ & $\ldots$ & $27.4$ \\
$u$       & $26.1$ & $\ldots$ & $\ldots$ & $27.4$ \\
$g$       & $27.4$ & $\ldots$ & $\ldots$ & $27.1$ \\
$r$        & $27.5$ &  $\ldots$ & $\ldots$ & $\ldots$ \\
$i$         & $26.8$ &  $\ldots$ & $\ldots$ & $\ldots$ \\
$z$        & $26.1$ &  $\ldots$ & $\ldots$ & $\ldots$ \\
$y$/$Y$ & $24.9$   & $24.0$  & $26.7$  & $\ldots$ \\
$J$         & $\ldots$ & $24.2$  & $26.9$  & $\ldots$ \\
$H$        & $\ldots$ & $23.9$  & $26.0$  & $\ldots$ \\
$K$        & $\ldots$ & $\ldots$ & $25.8$  & $\ldots$ \\ 
\hline
\end{tabular}
\end{center}
\end{table}

\subsubsection{Catalog Demographics}\label{sssec:exp_cat_demo}

In the top panel of Figure \ref{fig:cats_m} we show the cumulative distribution of true apparent magnitudes in our simulated catalog. The line style switches from solid to dashed when the distribution passes the $5{\sigma}$ limiting magnitude for detection in the filter. From this, we can see that for the LSST photometry, ${\sim}25\%$ of the galaxies in our sample remain undetected in the $u$-band, but all galaxies are detected in $grz$-bands, and almost all in $y$-band. All galaxies are detected in the $i$-band by design\footnote{However, we note that this does assume a perfect observing efficiency and in reality some small population of low surface-brightness galaxies that have an integrated flux brighter than $i\sim25$ mag might be undetected.}, since we impose a cut\footnote{Recall that the $i<25$ mag cut is imposed in order to compare the photo-$z$ results with the LSST science requirements, as described in Section \ref{sec:intro}.} of $i<25$ mag, which is brighter than the $5\sigma$ detection limit of $26.8$ mag.

\begin{figure}
\begin{center}
\includegraphics[width=8.5cm,trim={0cm 0cm 0cm 0cm}, clip]{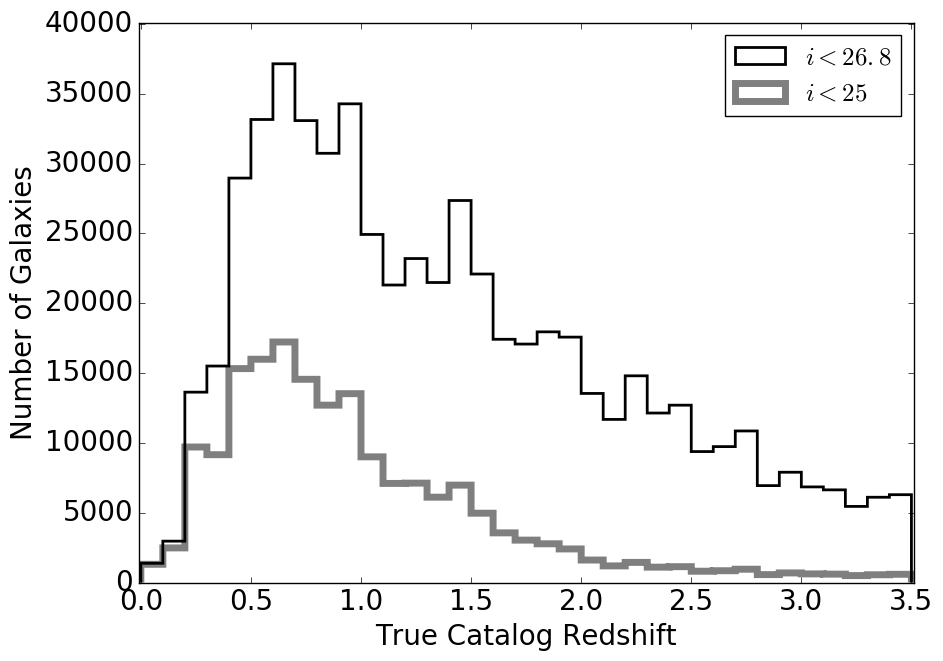}
\includegraphics[width=8.5cm,trim={0cm 0cm 0cm 0cm}, clip]{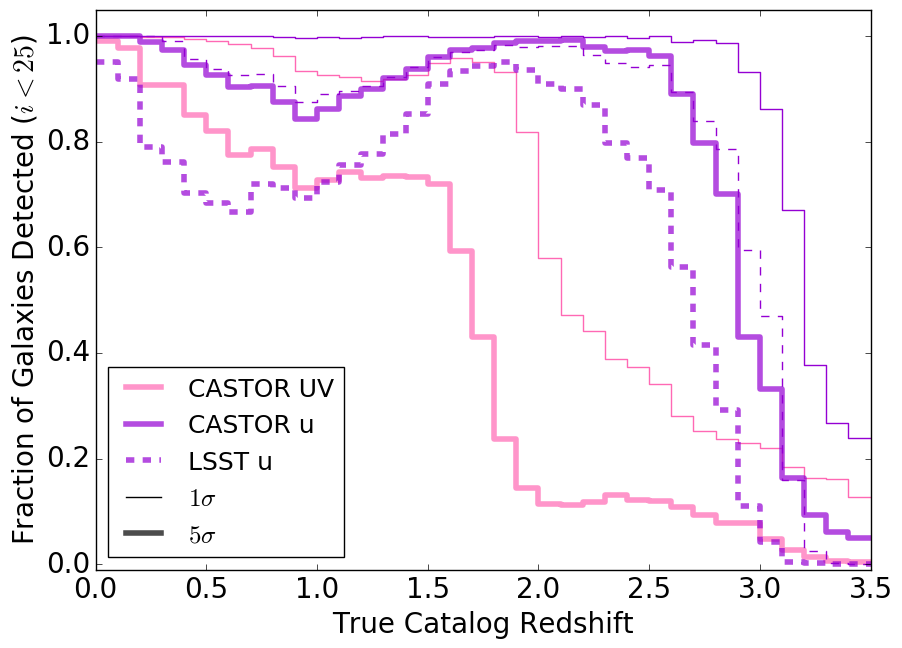}
\includegraphics[width=8.5cm,trim={0cm 0cm 0cm 0cm}, clip]{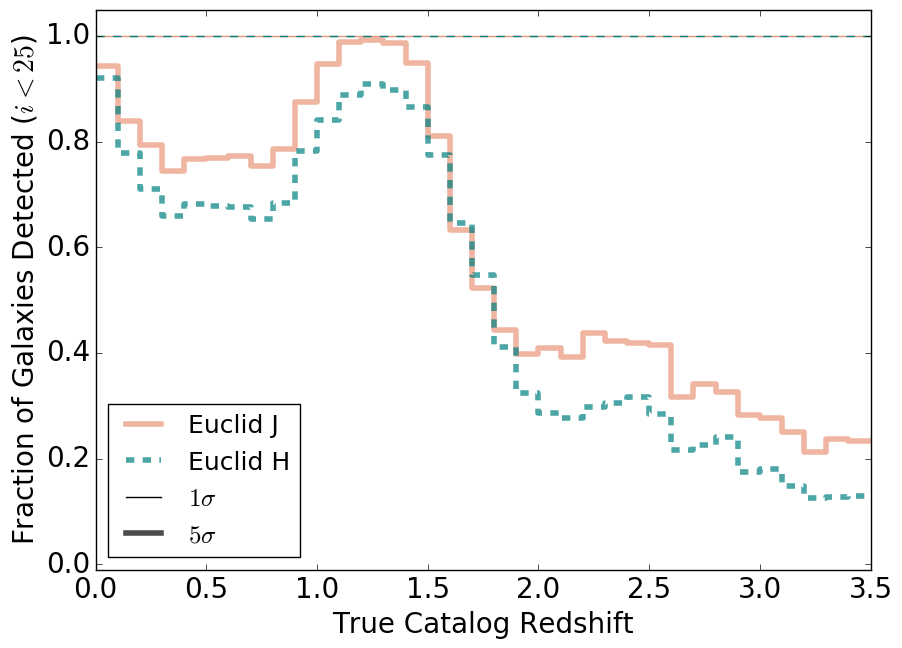}
\caption{{\it Top:} The simulated galaxy catalog's redshift distribution for galaxies with $i<25$ mag (thick grey) and $i<26.8$ mag (thin black), the LSST's 10-year $5\sigma$ and $1{\sigma}$ limiting magnitudes. {\it Middle:} Fraction of $i<25$ mag galaxies detected in the CASTOR $UV$ (pink) and $u_c$ (purple) bandpasses compared to the LSST $u$-band (dashed purple), for $1$ and $5{\sigma}$ detection limits (thin and thick lines, respectively). {\it Bottom:} Fraction of $i<25$ mag galaxies detected in the Euclid $J$ (solid salmon) and $H$ (dashed teal) filters, for $1$ and $5{\sigma}$ detection limits (thin and thick lines, respectively). Note that the thin lines are at $y=1$ because catalog galaxies with $i<25$ mag are completely detected by Euclid at $1{\sigma}$. \label{fig:cats_z}}
\end{center}
\end{figure}

In Figure \ref{fig:cats_m} we have included the Euclid near-infrared photometry in our plots of true catalog magnitude distributions (top panel) and error {\it vs.} magnitude (bottom panel). We can see that the cumulative distribution of true magnitudes are very similar for Euclid-$Y$ and LSST-$y$ (top panel), but that the 10-year LSST photometry has significantly smaller errors than Euclid's (bottom panel). Our treatment of these two $y$-bands in our photo-$z$ estimates is discussed in Section \ref{sssec:exp_cat_yY}. The photometric quality offered by WFIRST is also illustrated in the lower panel of Figure \ref{fig:cats_m}. The WFIRST limiting magnitudes are sufficiently deep that all catalog galaxies are detected in all four WFIRST filters.

In Figure \ref{fig:cats_z} we show the redshift distributions of the simulated galaxy catalog. Since we use each catalog galaxy as the ``test" galaxy in turn, the ``test" and ``training" sets have the same redshift distributions. This similarity helps to avoid redshift bias in the photo-$z$ results, but is not a likely property of future spectroscopic training sets, as discussed below. In the top panel of Figure \ref{fig:cats_z} we show the true redshift distribution of our simulated catalog, with applied cuts of $i\leq25$ and $\leq26.8$ mag (most of our simulations use the former).

In the middle panel of Figure \ref{fig:cats_z} we show the fraction of catalog galaxies detected in the CASTOR $u_c$ and $UV$ passbands as a function of true redshift, compared to LSST $u$, for detection limits of $1{\sigma}$ and $5{\sigma}$. We can see that the fraction of galaxies detected in the CASTOR $u$ passband at $5{\sigma}$ is approximately equivalent to that detected by the LSST $u$ filter at $1{\sigma}$ (thick solid and thin dashed purple lines, respectively). The sharp drop in fraction of galaxies detected in CASTOR $UV$ at $5{\sigma}$ indicates that adding CASTOR $UV$ photometry will probably only be able to improve photometric redshift estimates for $z\lesssim1.7$. 

In the bottom panel of Figure \ref{fig:cats_z} we show the fraction of catalog galaxies detected in the Euclid $J$ and $H$ filters as a function of true redshift, for detection limits of $1{\sigma}$ and $5{\sigma}$. Like the CASTOR $UV$ passband, the impact of including Euclid $5{\sigma}$ photometry in the photo-$z$ estimates will drop for $z\gtrsim1.7$. WFIRST detection fractions are not shown in any panel because, as with Euclid $1{\sigma}$ photometry, all LSST galaxies with $i<25$ magnitude are completely detected in the four WFIRST filters.

Throughout most of this work we will use a training set that has the same distributions of redshift and magnitude as the test set, and we emphasize that this is an ideal case (but appropriate for this work because our focus is on the impact of photometry, exclusively, on the photo-$z$ results). In reality, spectroscopic catalogs have some inherent bias and are not perfectly representative of the test set (which is an extremely difficult data product to build; e.g., \citealt{2015APh....63...81N}). In some of our simulations we consider a training set that has better or deeper photometry than a test set from a wide-field imaging campaign, and in some cases we also change the limiting magnitudes such that the test and training sets are no longer matched -- but only when the proposed surveys seem likely to provide such a data product.

\subsubsection{LSST $y$ {\it vs.} Euclid $Y$}\label{sssec:exp_cat_yY}

\begin{figure}
\begin{center}
\includegraphics[width=8.5cm,trim={0cm 0cm 0cm 0cm}, clip]{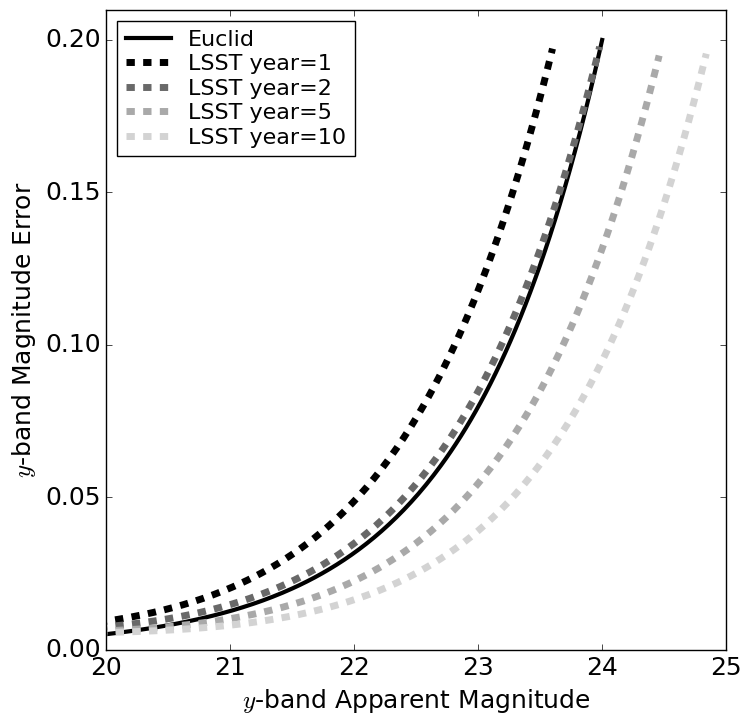}
\caption{Magnitude error {\it vs.} apparent magnitude for the Euclid $Y$-band (black solid line) and the LSST $y$-band at years 1, 2, 5, and 10 (black to light grey dashed lines). All lines cut off at the $5{\sigma}$ detection limit. It is evident that by year 2 of LSST, its $y$-band photometry will have a better accuracy than the Euclid $Y$-band. \label{fig:ylims}}
\end{center}
\end{figure}

The LSST $y$- and Euclid $Y$-band filters will not have exactly the same transmission function (the Euclid $Y$-band filter will be somewhat shifted redward and have sharper edges\footnote{As shown in the bottom-middle figure at \url{https://www.euclid-ec.org/?page\_id=2490}.}), but for the purposes of this work we consider the two filters to contribute equivalent information about a galaxy's SED to our photo-$z$ estimator, and use only one or the other. This is the most appropriate choice for our particular photo-$z$ estimator, because including colors $z-y$, $z-Y$, $y-J$, and $Y-J$ when calculating the Mahalanobis distance in color-space would give that region of the SED twice as much influence in Equation \ref{eq:DM}. We leave a study of whether two nearly-superimposed filters could provide additional SED information for e.g., emission line galaxies, for future work.

In Figure \ref{fig:ylims} we plot the apparent magnitude error {\it vs.} apparent magnitude for Euclid $Y$-band and the LSST $y$-band at survey years 1, 2, 5, and 10. It is clear that LSST's $y$-band photometric quality will be approximately equivalent to Euclid's $Y$-band with the LSST 2-year data release. In this work, we do not simulate the LSST photo-$z$ results at phases $<$2 years, so in all cases our simulations use the LSST $y$-band instead of the Euclid $Y$-band.

As a final note, the $y$-band filter of any ground-based facility like LSST will be impacted by a variable water absorption feature. However, it should be possible to correct the $y$-band colors to a fiducial filter system to within acceptable uncertainties, given a measure of the precipitable water vapor and a rough SED shape determined from the other five LSST bands (S.~Schmidt, private communication). Thus, we have not explicitly included the effects of water vapor on the $y$-band photometry in this work.

\subsection{Evaluation Methodology}\label{ssec:exp_stat}

To the results of our simulations we apply a similar evaluation methodology as used in \citet{2018AJ....155....1G}. Since we are using the CMNN estimator to predict the relative improvement in photo-$z$ due to changes in the input photometry (and not to predict absolute qualities), we do not extend our analysis to include predictions for cosmological parameters, but suggest readers consult \citet{2012MNRAS.420.3240N} and/or \citet{2018ARA&A..56..393M} for discussions of how photo-$z$ errors impact cosmological analyses. Below, we describe first our statistical measures of photo-$z$ quality, and then the common plot styles that we will use to represent our results in this work.

\subsubsection{Statistical Measures}\label{sssec:exp_stat_sm}

In our statistical analysis we use the photo-$z$ error $\Delta z_{(1+z)} = (z_{\rm true} - z_{\rm phot})/(1+z_{\rm phot})$, where $z_{\rm true}$ is the ``true" catalog redshift and $z_{\rm phot}$ is the photo-$z$. Including a factor of $(1+z)$ in the denominator compensates for larger uncertainties at high-$z$, and provides a comparable fractional error across our redshift range. For all of our results we calculate the robust standard deviation in $\Delta z_{(1+z)}$ as the FWHM of the interquartile range (IQR) divided by $1.349$ ($\sigma_{\rm IQR}$) and the robust bias as the mean value of $\Delta z_{(1+z)}$ in the IQR ($\overline{\Delta z_{\rm(1+z), IQR}}$). We bootstrap our uncertainties on these statistical measures by randomly drawing galaxy subsets and recalculating the statistics 1000 times, and then use the standard deviation of all recalculations as the error in the statistical measure. 

Outlier galaxies are identified as those with $\Delta z_{(1+z)} > 3\sigma_{\rm IQR}$ or $\Delta z_{(1+z)} > 0.06$, whichever is {\it larger}, where $\sigma_{\rm IQR}$ is calculated from all galaxies in $0.3 \leq z_{\rm phot} \leq 3.0$, i.e., outliers are defined globally. Since $\sigma$ changes with redshift outliers could instead be defined locally, as is done in other photo-$z$ analyses; however, we have adopted this global definition because it is equivalent to the LSST SRD's $3{\sigma}$ outlier, as mentioned in Section \ref{sec:intro}. \emph{Catastrophic outliers} with $|z_{\rm phot}-z_{\rm spec}| \geq 2$ are removed from the calculation of standard deviation and bias to keep these two statistical measures as representative of galaxies with ``good" photo-$z$ estimates. This removal is appropriate because catastrophic outliers are generally recognizable by the uncertainty in their photo-$z$ error returned by the CMNN estimator (and their photometric colors), and could be removed or flagged in a sample of real LSST photometric redshifts as well. Catastrophic outliers are included in the fraction of outliers statistical measure, though. 

Generally we discuss our results in three main redshift ranges: low ($0.3<z<0.6$), intermediate ($z\sim1$), and high ($z>2$). We do not include galaxies with $z<0.3$ in our statistical evaluations, as mentioned in Section \ref{sec:intro}, but also -- for interest's sake -- do not exclude the low-$z$ galaxies from our plots.

\subsubsection{Plot Styles}

To visualize our photo-$z$ results we create plots that compare the true $vs.$ photometric redshifts. We use a 2-dimensional histogram with log-normalized shading (such that black is always the most populated bin) in the densely sampled areas of $z_{\rm true}$ {\it vs.} $z_{\rm phot}$ space, and over-plot outlier galaxies with transparent red dots (see, e.g., Figure \ref{fig:NIRcomp_tzpz}). We also draw a solid line representing $z_{\rm true} = z_{\rm phot}$ to guide the eye. These plots are useful to obtain a global sense of the photo-$z$ quality and the structure in the outliers positions, especially the features that are perpendicular to the $z_{\rm true} = z_{\rm phot}$ which represent photo-$z$ degeneracies caused by the Balmer break passing between filters. 

To directly compare the statistical measures of photo-$z$ quality for our different simulations, we generate plots of the robust standard deviation, robust bias, and fraction of outliers as a function of binned $z_{\rm phot}$ (see, e.g., Figure \ref{fig:NIRcomp_stats}). We use a bin width of $0.3$ and a bin spacing of $0.15$ in $z_{\rm phot}$, such that bins overlap. Small vertical error bars mark the error in the statistical measurements, and wide horizontal bars mark the value of the statistic over the full range of $0.3 \leq z_{\rm phot} \leq 3.0$. Dashed lines are often used to represent the SRD's target values in these plots.

\section{ESA's Euclid Space Telescope}\label{sec:euclid}

\begin{table}
\begin{center}
\caption{Anticipated Progression of LSST and Euclid}
\label{tab:Euclid_LSST}
\begin{tabular}{ccccc}
\hline
\hline
Calendar & \multicolumn{2}{c}{LSST Main Survey} & \multicolumn{2}{c}{Euclid-Wide Survey} \\
Year & Year & Depth ($i$, $5{\sigma}$) & Year & Area Covered \\
 & & [mag] & & [$\rm deg^2$] \\
\hline
2024 & 1  & 25.6 & 3 & ${\sim}$2300 \\
2025 & 2  & 26.2 & 4 & ${\sim}$4300 \\
2027 & 5  & 26.5 & 7 & ${\sim}$7200 \\
2032 & 10 & 26.8 & 7 & ${\sim}$7200 \\
\hline
\end{tabular}
\end{center}
\end{table}

Euclid's launch is planned to be in 2021, and LSST is set to start operations in late 2022. In 2024, the 1-year LSST photometry will be available to be combined with the 3-year Euclid data set, the 2-year LSST data with the 4-year Euclid data, and so forth. Assuming that the LSST adopts a survey which uniformly progresses the depth of each annual data release for the Wide-Fast-Deep Main Survey of $18,000$ square degrees, we list the $5{\sigma}$ $i$-band limiting magnitude at survey years 1, 2, 5 and 10 in Table \ref{tab:Euclid_LSST}. Euclid will adopt a step-and-stare strategy which surveys a new area to full depth every year, as illustrated in Figure 6.11 of \citet{2011arXiv1110.3193L}. We list the Euclid-Wide survey area which will be complete after 3, 4 and 7 years in Table \ref{tab:Euclid_LSST}. Once the Euclid-Wide survey is complete, it will have an overlap area of ${\sim}40\%$ of the LSST's $18,000$ square degree main survey \citep{2017ApJS..233...21R}. Since \citet{2018AJ....155....1G} demonstrated that LSST $y$-band data primarily helps improve the photo-$z$ quality for galaxies with $1<z_{\rm phot}<1.5$, we expect the addition of Euclid data to improve the results for at least $z_{\rm phot}>1$.

\begin{figure}
\begin{center}
\includegraphics[width=7cm]{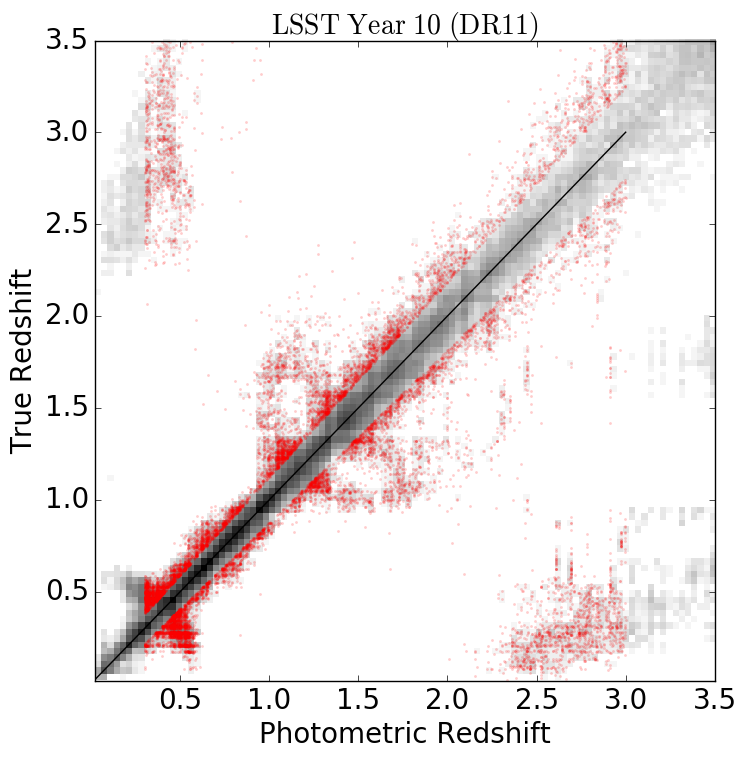}
\includegraphics[width=7cm]{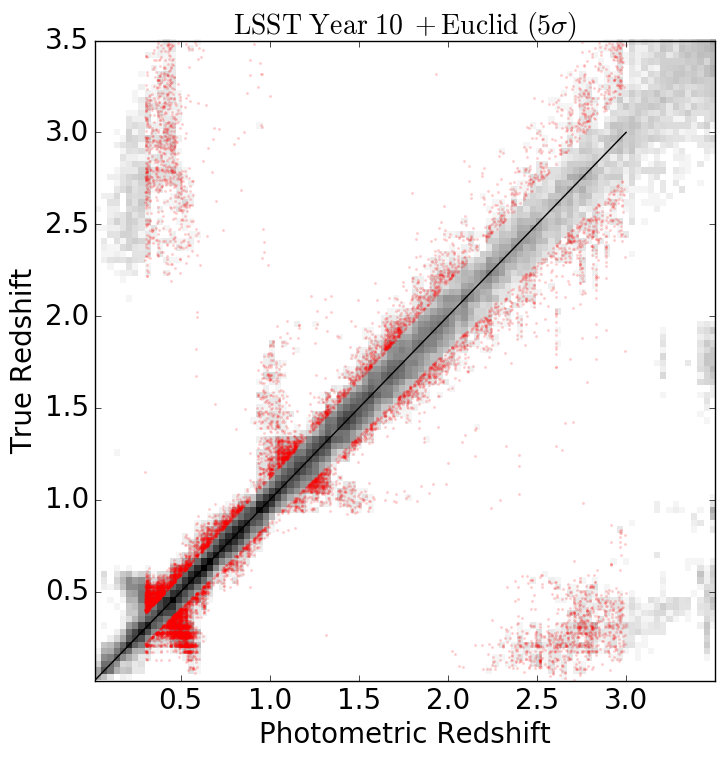}
\includegraphics[width=7cm]{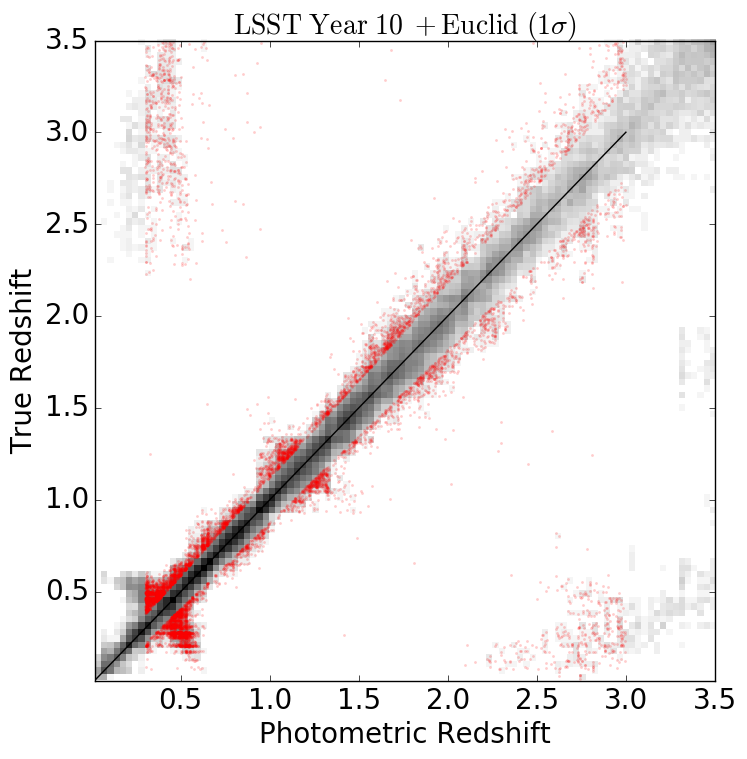}
\caption{True $vs.$ photometric redshifts when photometry from a 10-year LSST survey is used alone (top), and when Euclid NIR {\it JH} filters' $5{\sigma}$ and $1{\sigma}$ photometry are included (middle and bottom, respectively). The 2-dimensional histogram's scale is log-normalized and the most populous bin is black. Red points are outlier galaxies. \label{fig:NIRcomp_tzpz}}
\end{center}
\end{figure}

\begin{figure*}
\begin{center}
\includegraphics[width=7.5cm]{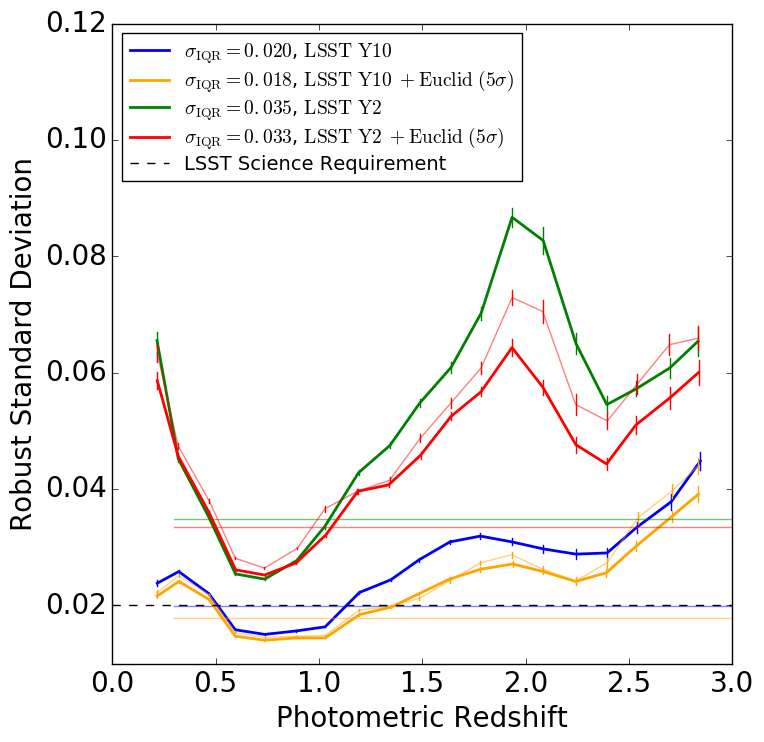}
\includegraphics[width=7.5cm]{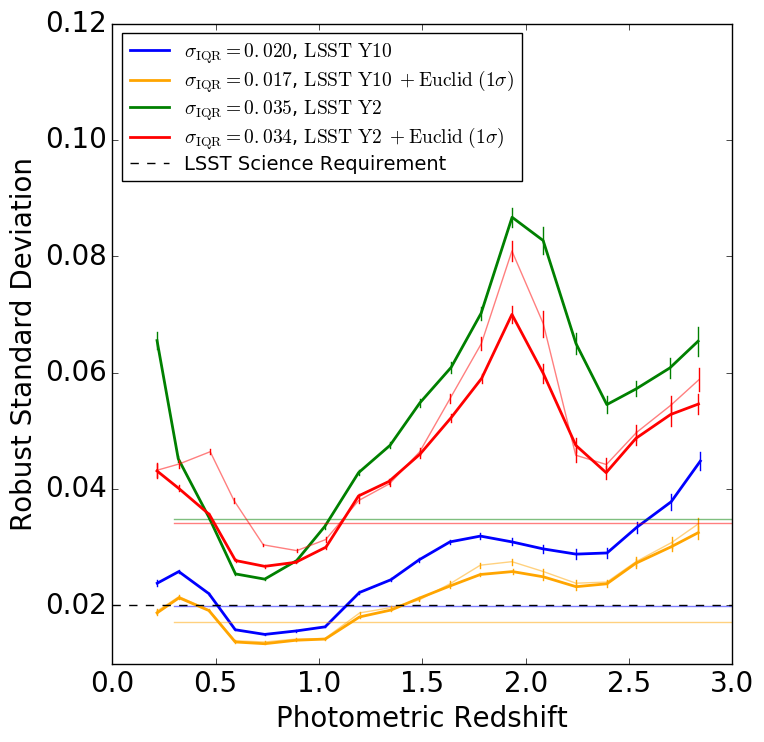}
\includegraphics[width=7.5cm]{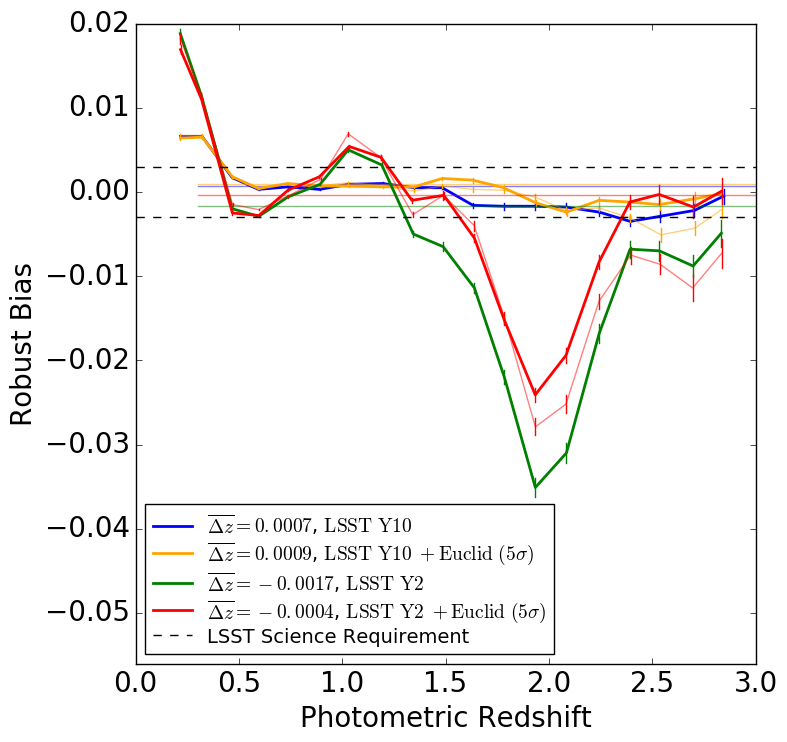}
\includegraphics[width=7.5cm]{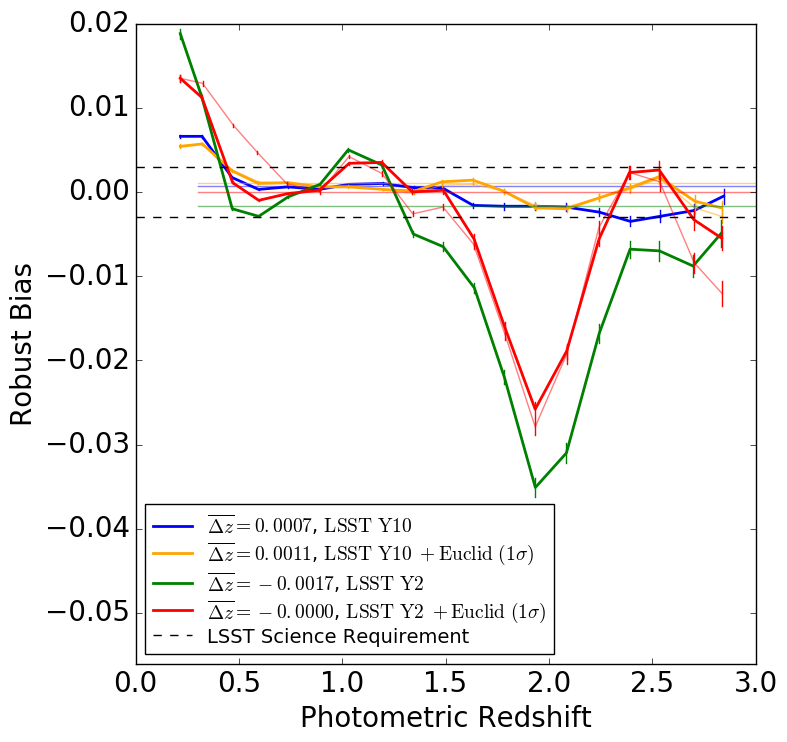}
\includegraphics[width=7.5cm]{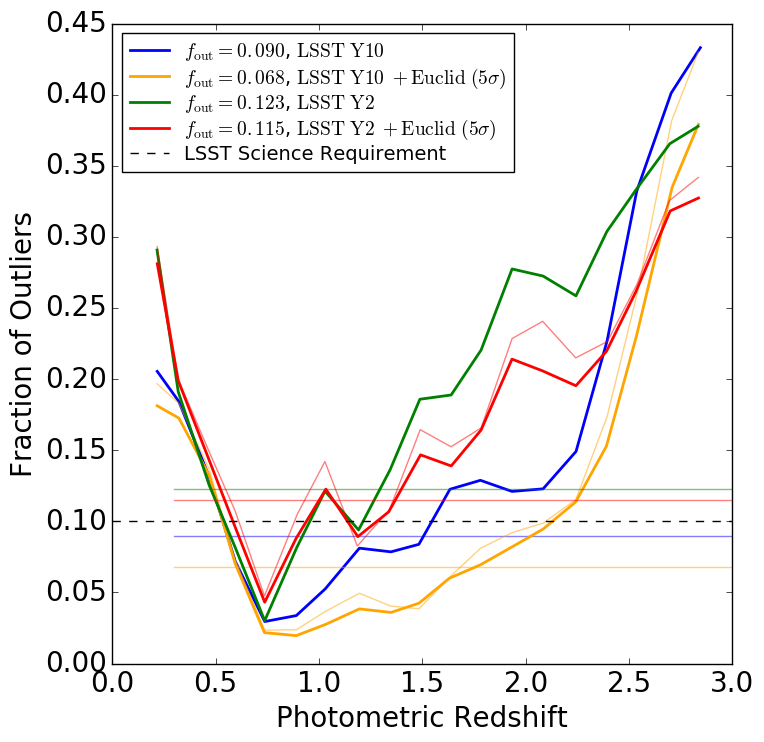}
\includegraphics[width=7.5cm]{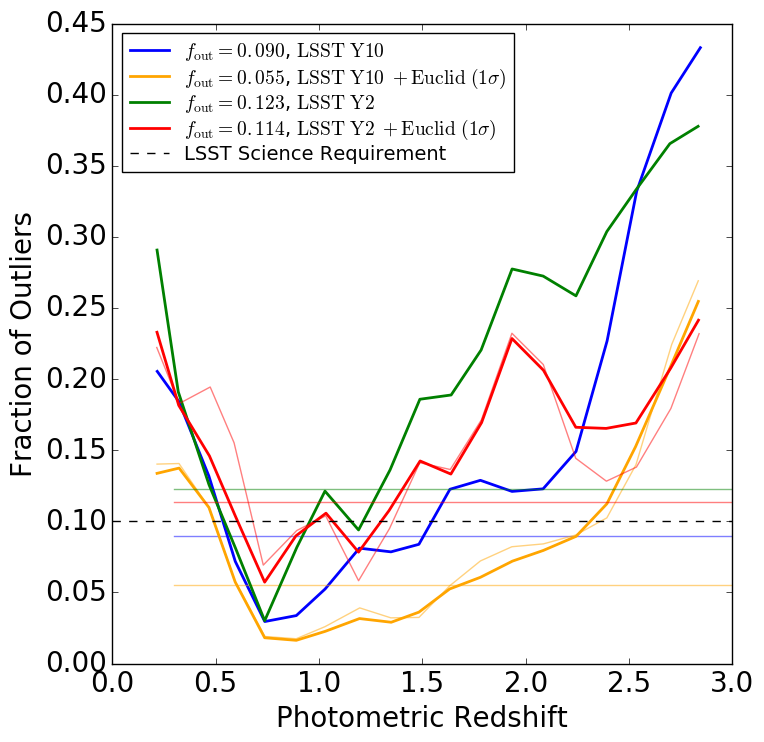}
\caption{Statistical measures for LSST photo-$z$ results show the impact of including Euclid NIR photometry with $5{\sigma}$ and $1{\sigma}$ detection limits (left and right columns respectively). {\it From top to bottom}: the robust standard deviation, robust bias, and fraction of outliers in bins of $z_{\rm phot}$. The green and blue lines represent photo-$z$ results based only on LSST photometry with $5{\sigma}$ limiting magnitudes equivalent to a baseline survey strategy at 2 and 10 years, respectively. The red and orange lines represent the photo-$z$ results at LSST years 2 and 10 when Euclid {\it JH} photometry is added (thin lines are the results when NIR photometry is included for \emph{all} galaxies even if does not help the photo-$z$ estimate, as described in the text). Horizontal solid lines represent the statistics over the full range of $0.3 \leq z_{\rm phot} \leq 3.0$, and dashed lines show the LSST's SRD target values as a reference point. \label{fig:NIRcomp_stats}}
\end{center}
\end{figure*}

We estimate photometric redshifts for simulated galaxies within the 7-year Euclid footprint and evaluate how the impact of adding Euclid photometry changes as the LSST survey adds depth. For the training set we assume that a field of galaxies with spectroscopic redshifts has been observed by both surveys. We simulate the observed apparent magnitudes and their uncertainties using the LSST 10-year $5{\sigma}$ limiting magnitudes and Euclid's $5{\sigma}$ limiting magnitudes for {\it JH}, as listed in Table \ref{tab:maglims}. We apply detection threshold cuts in observed apparent magnitude of $i<25$ mag, the LSST 10-year $5{\sigma}$ limiting magnitude for $ugrzy$, and Euclid's $5{\sigma}$ limiting magnitudes for {\it JH}.

For the test sets, we simulate the observed apparent magnitudes and their uncertainties using the 2- and 10-year $5{\sigma}$ LSST limiting magnitudes and the Euclid $5{\sigma}$ limiting magnitudes. The LSST's 2-year depths are $25.2$, $26.5$, $26.7$, $25.9$, $25.2$, $24.0$ mag in filters {\it ugrizy}, respectively. We apply detection threshold cuts in observed apparent magnitude of $i<25$ mag, the LSST 2- or 10-year $5{\sigma}$ limiting magnitude for $ugrzy$, and Euclid's $5{\sigma}$ or $1{\sigma}$ limiting magnitudes for {\it JH}. The Euclid $1{\sigma}$ limits are $25.95$ and $25.65$ mag in $J$ and $H$, respectively. Including galaxies detected at $1{\sigma}$ is the best way to incorporate Euclid ``non-detections" with the CMNN estimator, because there is not a simple way to include upper limits in the Mahalanobis distance (Section \ref{ssec:exp_cmnn})\footnote{One alternative method to include non-detections in the CMNN photo-$z$ estimates -- which we do not explore in this work -- is to use them as a prior or a weight when constructing the CMNN subset of training galaxies.}. Recall from Figure \ref{fig:cats_z} that our set of simulated galaxies with $i<25$ mag is completely detected in the Euclid NIR filters at $1{\sigma}$, but $>$50\% incomplete at redshifts $\gtrsim1.7$ with a $5{\sigma}$ cut.

In Figure \ref{fig:NIRcomp_tzpz} we plot the true {\it vs.} photometric redshift results for a 10-year LSST survey alone, and when Euclid detections at $5{\sigma}$ and $1{\sigma}$ are included. These plots show how the addition of Euclid data reduces the number of outliers (as defined in Section \ref{sssec:exp_stat_sm}), especially the catastrophic outliers with over-estimated photo-$z$ (the cloud of red points in the lower-right corner). The fact that Euclid also reduces the scatter around $z_{\rm true}=z_{\rm phot}$ for non-outlier galaxies is almost imperceptible in these plots, but in Figure \ref{fig:NIRcomp_stats} we show the standard deviation, bias, and fraction of outliers as a function of binned photo-$z$ for LSST year 2 and 10, with and without the addition of $5{\sigma}$ or $1{\sigma}$ Euclid photometry. When considering Figure \ref{fig:NIRcomp_stats} it is important to remember that the definition of an outlier depends on the standard deviation and includes catastrophic outliers ($|z_{\rm true}-z_{\rm phot}|\geq2$), whereas the robust standard deviation and bias excludes catastrophic outliers. This leads to the standard deviation and fraction of outliers being inversely correlated in some of our results.

\begin{figure}
\begin{center}
\includegraphics[width=8.5cm]{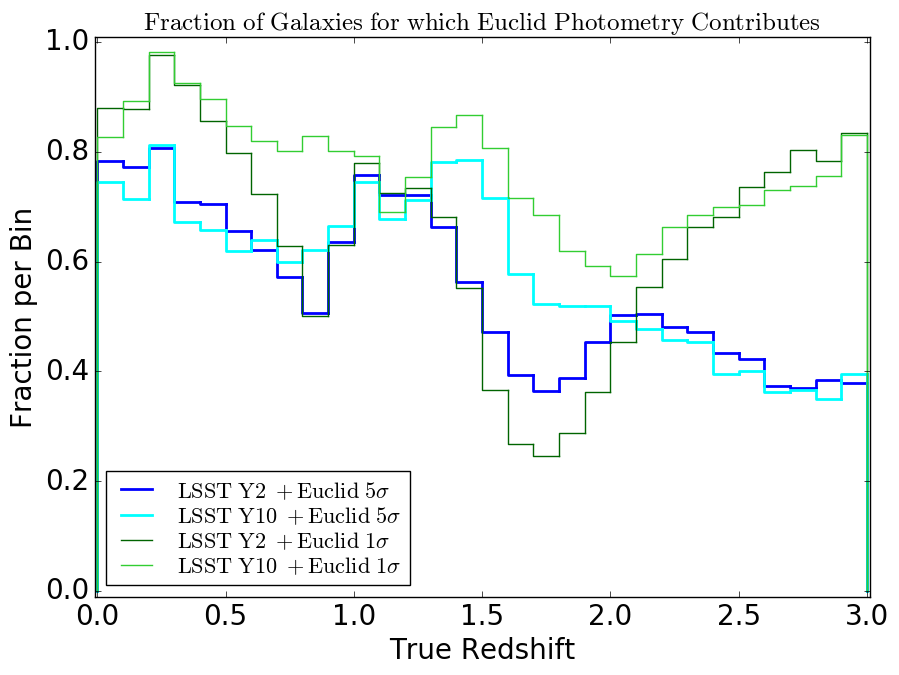}
\includegraphics[width=8.5cm]{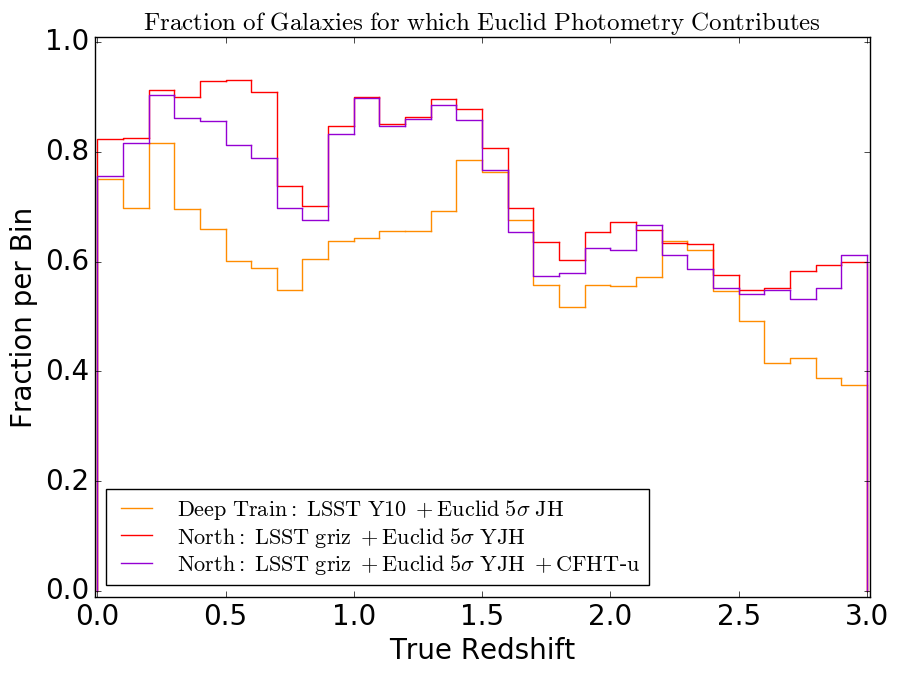}
\caption{The fraction of galaxies in each redshift bin that were detected by Euclid \emph{and} for which Euclid NIR photometry was deemed to be worth including in the photo-$z$ estimate for a given simulation (as labeled in the legend). Galaxies were considered detected by Euclid if they had $J<24.2$ or $J<25.95$ mag for the $5{\sigma}$ and $1\sigma$ detection limits (all simulations), and/or if they had $Y<24.0$ mag (North survey simulations).   \label{fig:frac_restrict}}
\end{center}
\end{figure}

With the CMNN photo-$z$ estimator there can be instances when additional filters increase the number of degrees of freedom in Equation \ref{eq:DM} but the photometric quality is insufficient to produce an improvement in the photo-$z$ results (as discussed in Section \ref{ssec:exp_cmnn}). We find that the addition of Euclid data to the 2-year LSST photometry is one of those cases. In Figure \ref{fig:NIRcomp_stats} the thinner red lines represent the photo-$z$ quality when the Euclid photometry is added to the 2-year LSST photometry for all galaxies, and we see that this leads to a \emph{larger} standard deviation in the redshift bins $0.5<z_{\rm phot}<1$. The simplest way to mitigate this issue with the CMNN estimator is to not include additional photometry if it is unlikely to improve the photo-$z$ estimate. We explored our simulated data and found that the photo-$z$ error $\Delta z_{\rm 1+z}$ was likely to be larger (i.e., the photo-$z$ was less accurate) if (1) the photo-$z$ {\it uncertainty} was larger when NIR data was included and (2) the galaxy had a bluer color in the bluer filters. Based on this, we imposed a restriction that the NIR photometry be excluded from the photo-$z$ estimate for galaxies with $u-g<0.5$ or $g-r<0.4$ \emph{unless} it produced smaller a photo-$z$ uncertainty than the LSST optical filters alone. We find that with this restriction the apparent deterioration of the results is mitigated, as shown by the thicker red lines in Figure \ref{fig:NIRcomp_stats}. This restriction still allows the NIR photometry to contribute to the photo-$z$ estimates of most simulated galaxies, as shown in Figure \ref{fig:frac_restrict}. There remains a small deterioration in standard deviation in redshift bins $0.6<z_{\rm phot}<0.8$ (thick green {\it vs.} red line in the top right plot), which suggests that our restriction could be further optimized, but we do not make any more complicated restrictions on the addition of NIR photometry at this time. We find that this restriction is not as necessary for the 10-year LSST photometry (i.e., the thin and thick orange lines in Figure \ref{fig:NIRcomp_stats} show only minor differences).

Figure \ref{fig:NIRcomp_stats} demonstrates that incorporating the $5{\sigma}$ Euclid photometry will improve both the 2- and 10-year LSST photo-$z$ results at intermediate and high redshifts, $z_{\rm phot} > 1$. This improvement is larger and extends to higher redshifts when the Euclid detection limit is extended to $1{\sigma}$. This improvement was expected because intermediate redshifts are where the $4000$ $\rm \AA$ Balmer break becomes redshifted beyond the optical filters. The NIR colors provide the location of the break, and prevent truly low-$z$ galaxies from being assigned a high-$z_{\rm phot}$. Figure \ref{fig:NIRcomp_stats} also shows that the $1{\sigma}$ Euclid photometry improves the results at low redshift, $z_{\rm phot} <0.5$, for the same reason.

Quantitatively, Figure \ref{fig:NIRcomp_stats} shows that including Euclid $5{\sigma}$ {\it JH} photometry could provide a ${\sim}20\%$ reduction in standard deviation (at LSST years 2 and 10) for redshifts $z_{\rm phot}>1$, and ${\sim}6\%$ (year 2) and ${\sim}10\%$ (year 10) reductions over the full redshift range ($0.3 < z_{\rm phot} < 3.0$). For the bias, Euclid offers a ${\sim}25\%$ reduction at LSST year 2 for $z_{\rm phot}>1$, but has little impact in the later years of the survey. Most remarkably, Euclid would provide a ${\sim}25\%$ (year 2) and a ${\sim}40\%$ (year 10) reduction in the fraction of outliers for $z_{\rm phot} > 1$, and a ${\sim}6\%$ (year 2) and a ${\sim}25\%$ (year 10) reduction over the full redshift range. If Euclid $1{\sigma}$ detections are included, the reduction in the fraction of outliers increases to $\gtrsim50\%$ for $z_{\rm phot} > 1$, and ${\sim}40\%$ over the full redshift range for LSST year 10.

Given the clear and beneficial impact of Euclid photometry on LSST photo-$z$ estimates, the question arises of which has \emph{more} of an impact on the photo-$z$ estimates: adding Euclid photometry for more LSST galaxies as the overlap area increases over time, or improving the LSST {\it ugrizy} photometry with the yearly progression in imaging depth? With respect to the bias and standard deviation, we find that improving the optical depth typically has a greater impact. There is one notable exception to this: for LSST year 5 (not shown in our plots) the addition of Euclid $1{\sigma}$ photometry could bring the standard deviation down to the 10-year values in a limited range of redshift bins ($1.3 \lesssim z_{\rm phot} \lesssim 1.6$ and $z_{\rm phot} > 2.7$). With respect to the fraction of outliers, however, for LSST at year 5 we find that the addition of Euclid $1{\sigma}$ photometry brings the fraction of outliers down to the 10-year values for all galaxies with $z_{\rm phot}>1$. Generally, we find that reducing the outliers at low- and high-redshifts is better accomplished with Euclid than by obtaining deeper LSST imaging (e.g., as seen in the bottom plots of Figure \ref{fig:NIRcomp_stats}).

As a final note, we consider whether the photo-$z$ improvements offered by Euclid NIR photometry might instead be achievable by obtaining more images with LSST in just the $z$ and $y$ filters. We simulate LSST-only photo-$z$ results with deeper $z$ and $y$ photometry and find that in order to decrease the standard deviation over the full redshift range of $0.3<z_{\rm phot}<3$ to a value below that offered by combining LSST and Euclid $5{\sigma}$ photometry, the LSST would have to more than double the total amount of integration time in both $z$ and $y$. This would require another ${\sim}2$ years of survey to cover $15000$ square degrees in the two filters, and still would not fully replicate the benefits of Euclid NIR photometry: additional depth in LSST $z$ and $y$ mainly improves the standard deviation for $z_{\rm phot}<1.5$, whereas Euclid improves it in the higher-$z_{\rm phot}$ bins \emph{and} provides a more drastic reduction in the fraction of outliers.

\subsection{The Impact of a Deeper Training Set}\label{ssec:euclid_deeptrain}

\begin{figure}
\begin{center}
\includegraphics[width=7.5cm,trim={0cm 0cm 0cm 0cm}, clip]{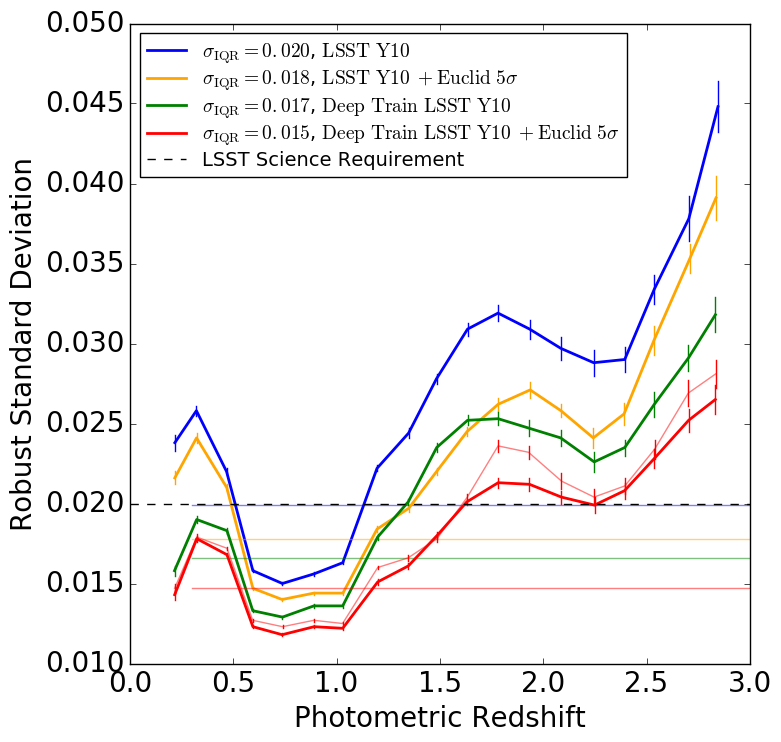}
\includegraphics[width=7.5cm,trim={0cm 0cm 0cm 0cm}, clip]{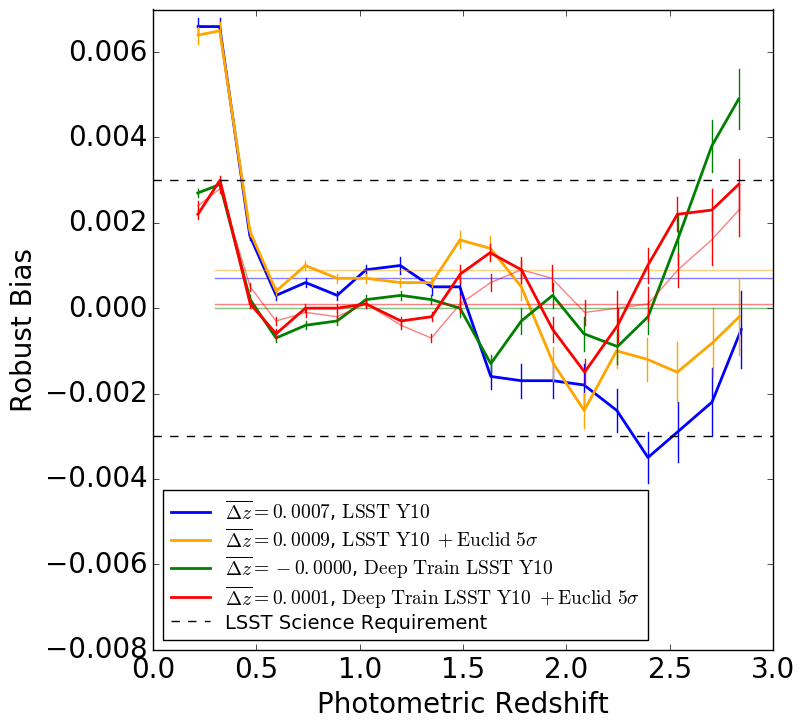}
\includegraphics[width=7.5cm,trim={0cm 0cm 0cm 0cm}, clip]{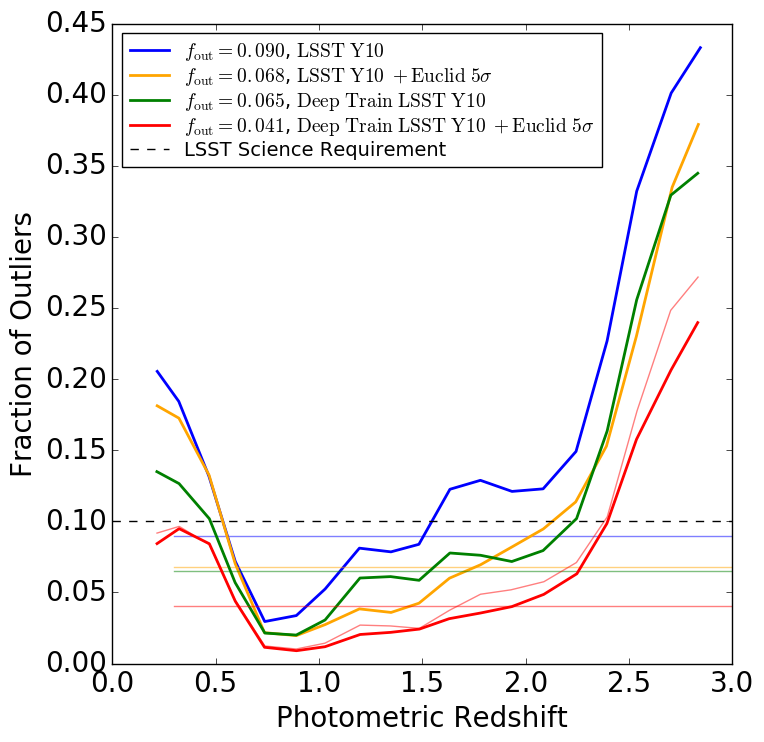}
\caption{Statistical measures of photo-$z$ results for a 10-year LSST survey without (blue) and with (orange) Euclid photometry included, and for a deeper training set derived from well-covered LSST (green) and Euclid areas (red). The thin and the horizontal lines have the same meaning as in Figure \ref{fig:NIRcomp_stats}. \label{fig:DT_stats}}
\end{center}
\end{figure}

Both the LSST and the Euclid surveys will include smaller areas in which a significantly larger number of images are obtained and deeper imaging stacks will be created. For the LSST deep drilling fields (DDFs), \citet{Ivezi__2019} describes a potential observing strategy that would obtain ${\sim} 25$ $30$-second consecutive visits in the {\it griz} filters every other night for four months in order to find high-$z$ supernovae. This would generate up to an {\it additional} $1500$ visits in {\it each} filter compared to the wide-fast-deep main survey. There will be at least 4 LSST DDF, totaling ${\sim}40$ $\rm deg^{2}$. Since \citet{2018AJ....155....1G} demonstrated that the $u$ and $y$-band filters are important for the quality of photometric redshifts, we assume that the observing strategy for any DDF to be used as a photo-$z$ spectroscopic training set would include, e.g., $500$ visits in each of filters $u$ and $y$, and $1000$ visits in each of filters {\it griz}. The $5{\sigma}$ limiting magnitudes of such a DDF deep stack would be $27.3$, $28.8$, $28.4$, $27.8$, $27.1$, and $25.5$ in filters {\it ugrizy}, respectively. Euclid is projected to have at least 2 deep fields that total ${\sim} 40$ $\rm deg^{2}$ in which the detection limit extends by ${\sim}2$ magnitudes (e.g., \citealt{2016ASPC..507..401J}). The $5{\sigma}$ limiting magnitudes of the Euclid deep fields would be $26.0$, $26.2$, and $25.9$ magnitudes in each of filters {\it YJH}.

We simulate a deep training set with apparent observed magnitudes using these $5{\sigma}$ limiting magnitudes, and apply cuts in apparent magnitude at these $5{\sigma}$ limits in all filters except $i$-band, where we retain the cut of $i<25$ mag because obtaining a representative set of spectroscopic redshifts for galaxies fainter than this is difficult and not guaranteed\footnote{Although with dedicated time on $10$ to $30$ m telescopes it might be possible to obtain a small, but very deep, training set, we leave further exploration of training set variations to future work and just consider a simple extension here.}. The $i<25$ mag restriction also keeps this deep training set approximately matched to the test set in terms of its $i$-band magnitude and redshift distribution, which helps to avoid introducing a bias in the CMNN photometric redshift results. The photometric quality of the training set  is thus improved in two ways: (1) the photometric errors and observational scatter are reduced and (2) more galaxies are detected in more filters. We use a test set of galaxies with simulated photometric quality equivalent to a $10$ year LSST survey and the main Euclid survey. For both test and training sets, LSST $y$-band filter is used instead of the Euclid $Y$-band.

Figure \ref{fig:DT_stats} demonstrates the impact of using a deep training set on the photo-$z$ results by plotting the statistical measures of standard deviation, bias, and fraction of outliers as a function of redshift. As in the previous section, thin lines represent the results when the NIR photometry is included for all galaxies without a restriction, and the fraction of galaxies for which the NIR photometry is included in the photo-$z$ estimates for this simulation is shown in the bottom panel of Figure \ref{fig:frac_restrict}. As expected, the standard deviation, bias, and fraction of outliers are all improved by the use of a deeper training set and the inclusion of Euclid photometry. The greatest impact is seen in the intermediate- to high-$z$ galaxy bins ($z_{\rm phot}>1$), where the standard deviation is improved by ${\sim}25\%$ with the use of a deep training set and then \emph{an additional} ${\sim}25\%$ by the inclusion of Euclid photometry -- but there is also significant improvement over the full redshift range of $0.3<z_{\rm phot}<3.0$. Note that the bias in the high-$z$ bins switches from a slight overestimate (bias $<0$) to a slight underestimate (bias $>0$) when a deep training set is used. The alleviation of overestimates with a deeper training set is likely because the CMNN subsets of training galaxies will contain less high-$z$ interlopers when the photometric quality is improved. 

To explore the potential impact of a deeper training set on the photo-$z$ results earlier in the LSST survey, we repeat the simulation with test sets of a photometric quality equivalent to $2$ and a $5$ years, with and without Euclid {\it JH} photometry. We find that the deeper training set alone (no Euclid included) mitigates the bias at low- and intermediate-redshifts at year $2$ and $5$ (the unmitigated bias can be seen in the middle plots of Figure \ref{fig:NIRcomp_stats}). The use of a deeper training set at earlier times also helps to lower the fraction of outliers, but not as much as the addition of Euclid data. The biggest impact of the deep training set at early times is in the standard deviation: the year $2$ results are significantly reduced across the full redshift range, approaching the year $5$ results for $z_{\rm phot} > 1.7$, and are {\it better} than the year $5$ results for $z_{\rm phot} < 0.6$. When we use {\it both} a deep training set {\it and} Euclid photometry, the photo-$z$ standard deviation at year $2$ ($5$) is equivalent to or better than the results at year $5$ ($10$) across the full redshift range ($0.3<z_{\rm phot}<3.0$).

\subsection{LSST+Euclid Results for a Shallow Northern Field}\label{ssec:euclid_north}

\begin{figure}
\begin{center}
\includegraphics[width=7.5cm,trim={0cm 0cm 0cm 0cm}, clip]{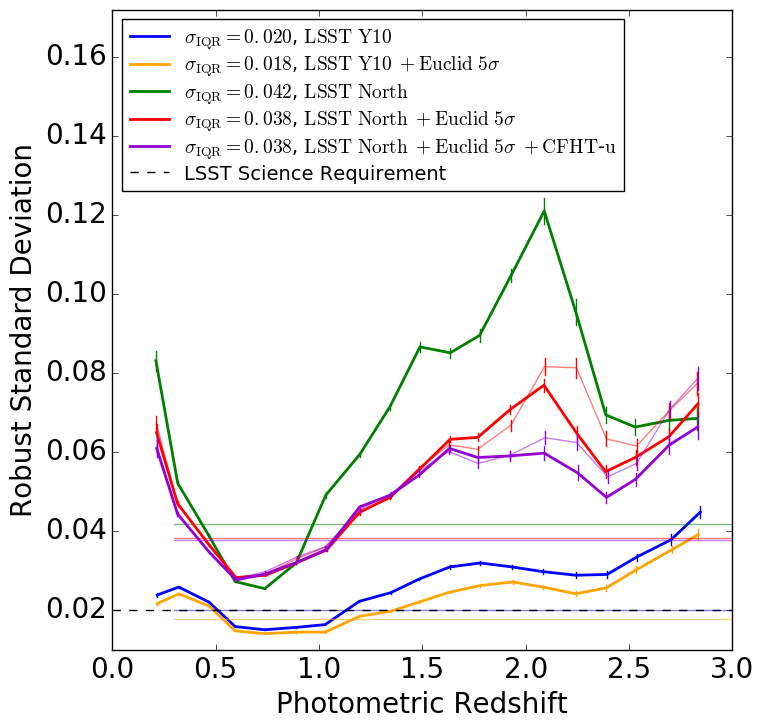}
\includegraphics[width=7.5cm,trim={0cm 0cm 0cm 0cm}, clip]{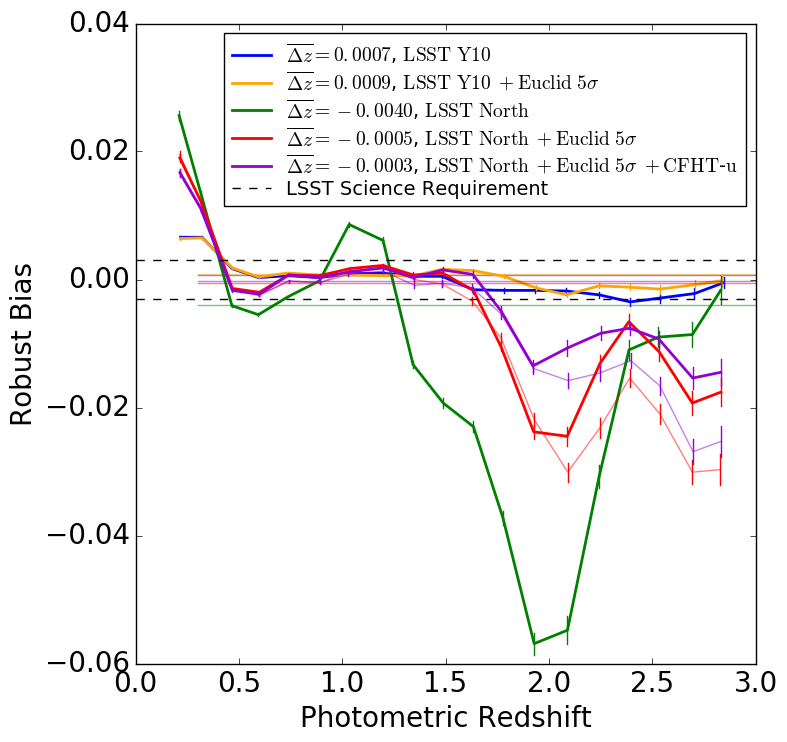}
\includegraphics[width=7.5cm,trim={0cm 0cm 0cm 0cm}, clip]{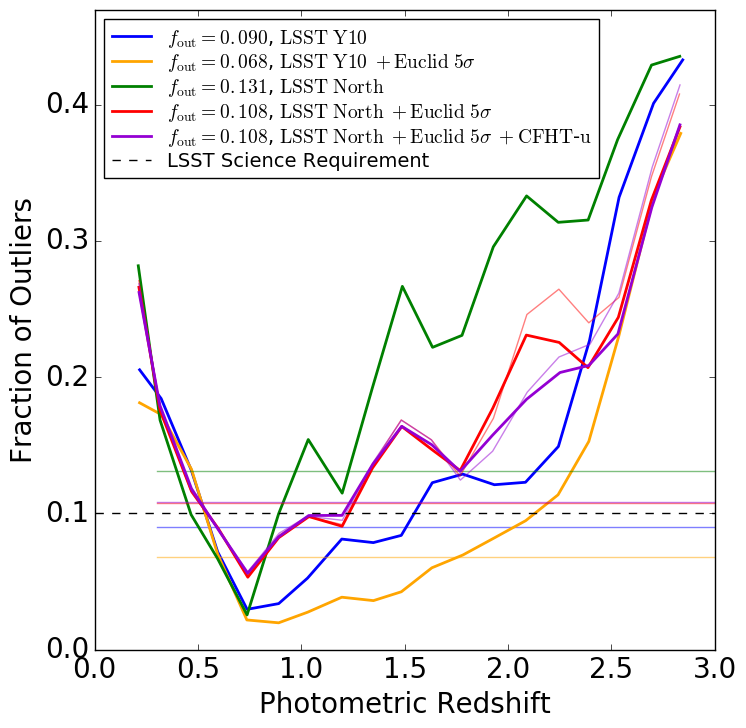}
\caption{Statistical measures of photo-$z$ results for a 10-year LSST survey without (blue) and with (orange) Euclid photometry included, compared to the results for a shallow northern LSST survey area (green) that overlaps with Euclid (red) and a CFHT $u$-band (purple) survey. The thin and the horizontal lines have the same meaning as in Figure \ref{fig:NIRcomp_stats}. \label{fig:SN_stats}}
\end{center}
\end{figure}

\citet{2017ApJS..233...21R} propose that LSST cover an additional northern survey area of ${\sim} 3000$ square degrees in order to increase the amount of overlap with Euclid. This proposed northern survey is shallower than the LSST main survey, with a total of 43 visits in filters {\it griz}. We test what the photo-$z$ results would be for this northern survey by simulating photometry with 6, 13, 13, and 11 visits in each of filters $g$, $r$, $i$, and $z$ respectively (i.e., distributing the 43 visits using the same relative fractions for each filter as the wide-fast-deep survey). The LSST $5{\sigma}$ detection limits in {\it griz} for this shallow survey would be $26.0$, $26.1$, $25.4$, and $24.6$ mag, respectively. This is significantly deeper than PanSTARRS, which goes to $23.3$, $23.2$, $23.1$, and $22.3$ mag in its {\it griz} filters \citep{2016arXiv161205560C}. For now we ignore the degradation of the photometry due to the fact that most of these visits will be at higher airmass (all photometric errors used in this work are for airmass 1.2, as described in Section \ref{sssec:exp_cat_err}). This is acceptable here because our main goal is to evaluate how NIR photometry affects the photo-$z$ in this shallow northern survey, and not the absolute quality of photo-$z$ results for such a survey. 

We simulate a test set with apparent observed magnitudes using the LSST $5{\sigma}$ limiting magnitudes quoted above and the Euclid $5{\sigma}$ limiting magnitudes for filters {\it YJH}. Note that Euclid $Y$-band is used here because there would be no LSST $y$-filter coverage in this shallow northern survey. We apply cuts in apparent magnitude of $i<25$ mag and at the $5{\sigma}$ detection limit for all other LSST and Euclid filters. For the training set we simulate photometry using the LSST 10-year and Euclid main survey $5{\sigma}$ limiting magnitudes, and apply cuts in apparent magnitude of $i<25$ mag and at the $5{\sigma}$ limits for the other filters. Note that in this simulation the training set is deeper than the test set, which could lead to a bias in the photo-$z$ results.

In Figure \ref{fig:SN_stats} we show the statistical measures of standard deviation, bias, and fraction of outliers for photometric redshifts based on the photometry of this proposed shallow northern survey (and for context we also show the LSST 10-year equivalent results). As in previous sections, the thin lines represent the results when the NIR photometry is included for \emph{all} galaxies, regardless of whether it might deteriorate the photo-$z$ estimate. We find that the shallow northern survey is similar in depth to the LSST year $1$ results (not shown in this work, but for reference see the year $2$ results in Figure \ref{fig:NIRcomp_stats}). Since plenty of science goals will be achievable with the LSST $1$ year data release photometric redshifts (e.g., as demonstrated by \citealt{2018AJ....155....1G}), the same applies to a shallow northern LSST survey area. As also seen for the LSST year 2 results, the photo-$z$ results for galaxies with $z_{\rm phot}>1$ are all improved when Euclid NIR is added to LSST photometry from a shallow northern survey. In particular, at redshifts $1<z<2.5$ the addition of Euclid photometry to a shallow northern LSST survey significantly improves the photo-$z$ results, decreasing the standard deviation, fraction of outliers, and absolute bias by up to ${\sim}30\%$ (Figure \ref{fig:NIRcomp_stats}).

Another option for this region will be to include photometry from the the Canada-France Imaging Survey (CFIS), which has been covering this area of the northern sky in the $u$-band with the Canada-France-Hawaii Telescope (CFHT) as a continuation of the Legacy for the U-band All-sky Universe (LUAU)\footnote{\url{http://www.cfht.hawaii.edu/Science/CFIS/}} program, for a wide variety of scientific applications. \citet{2017ApJ...848..128I} present the first results from the CFIS $u$-band component, which has a 5$\sigma$ limiting magnitude of ${\sim}24.4$ mag (i.e., approximately equal to 2 LSST $u$-band visits). To evaluate the impact of CFIS-$u$ coverage for this shallow northern survey, we simulated photo-$z$ results and include them in Figure \ref{fig:SN_stats}. We find a small decrease in the standard deviation in the lowest redshift bin ($z_{\rm phot}\sim0.5$) compared to using LSST and Euclid photometry only, and a more significant improvement at $z_{\rm phot}>1.7$ where the $u$-band data helps to resolve degeneracies between low- and high-$z$ galaxies. Similar results for the impact on photo-$z$ of adding CFHT $u$-band photometry to {\it grizy} survey data -- in this case from the Hyper Suprime-Cam Subaru Strategic Program (HSC-SSP) -- were presented by \citet{2019MNRAS.489.5202S}. They demonstrated that the photo-$z$ were most improved at $z<0.75$ and $z>2$, e.g., with ${\sim}25\%$ and ${\sim}40\%$ reductions in standard deviation, respectively.

As a final note, if this shallow northern survey was done as an \emph{extension} of the wide-fast-deep survey area (and not, e.g., as a separate mini-survey), then it would remove approximately $1.2$ visits per filter from all fields within the main survey's $18000$ square degrees. This is a small enough fraction that we do not simulate the overall impact of this loss of depth on the photo-$z$ for the LSST main survey.

\section{NASA's WFIRST Mission}\label{sec:wfirst}

\begin{figure}
\begin{center}
\includegraphics[width=7.5cm,trim={0cm 0cm 0cm 0cm}, clip]{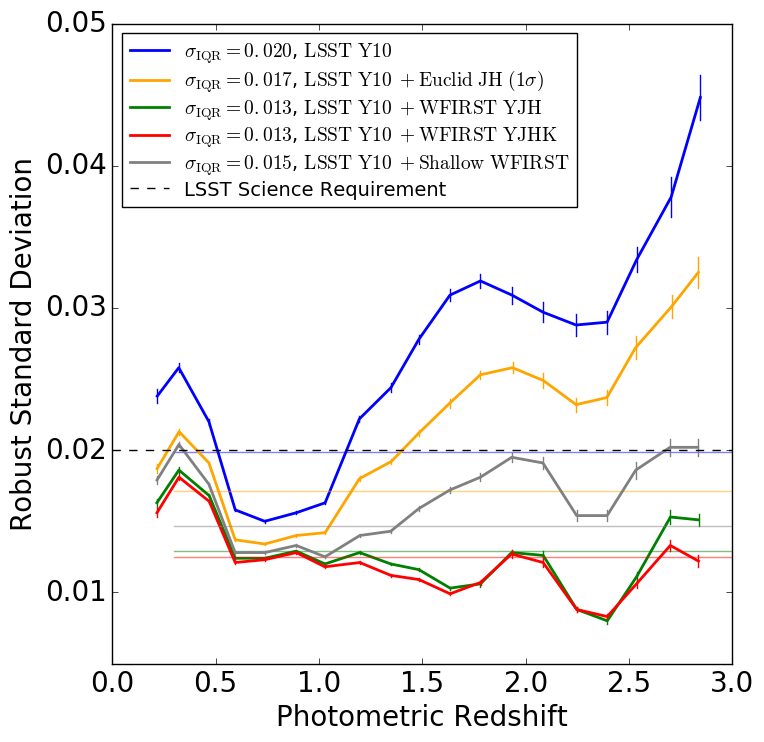}
\includegraphics[width=8cm,trim={0cm 0cm 0cm 0cm}, clip]{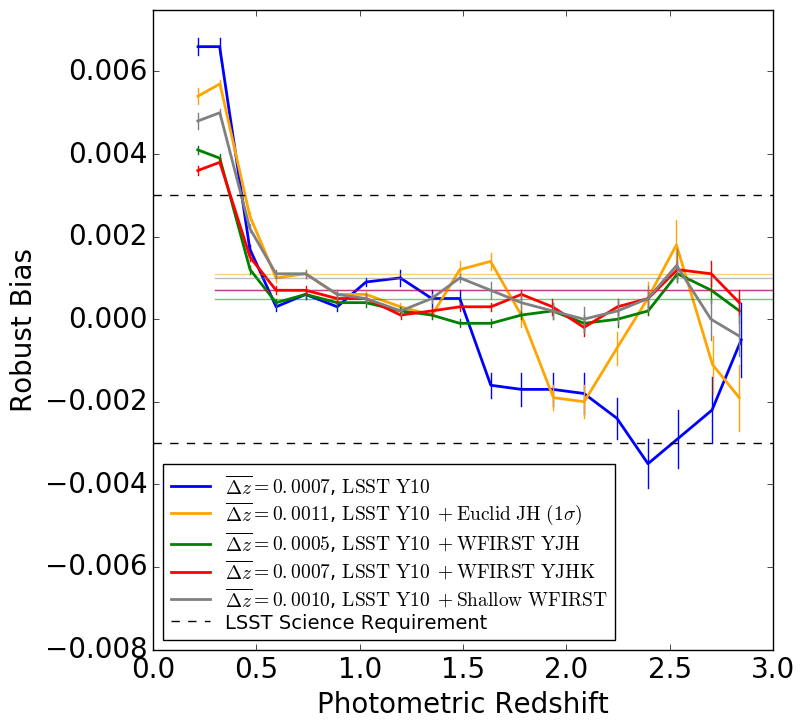}
\includegraphics[width=7.5cm,trim={0cm 0cm 0cm 0cm}, clip]{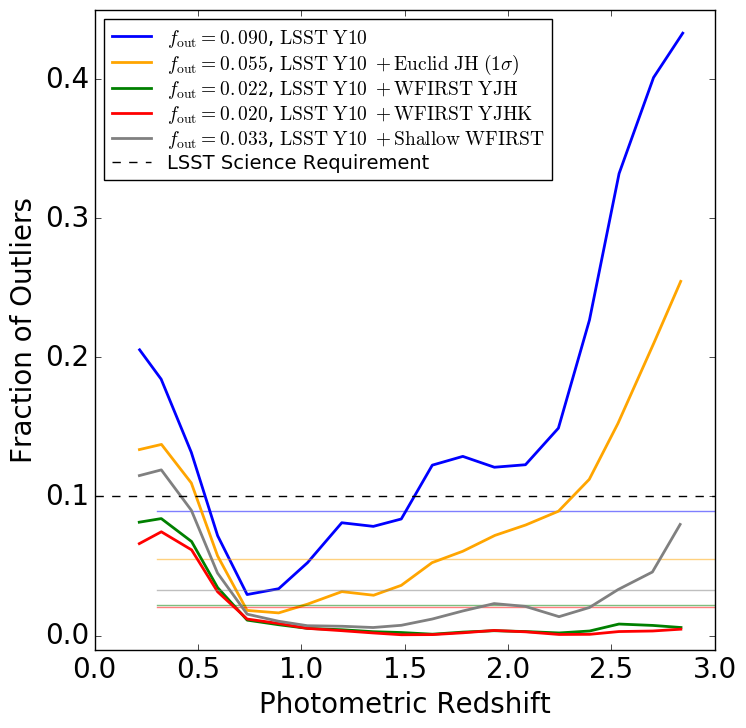}
\caption{The statistical measures of standard deviation (top), bias (middle), and fraction of outliers (bottom) for our photo-$z$ results with: LSST photometry only (blue); Euclid $1{\sigma}$ (orange; with the restriction described in Section \ref{sec:euclid}); WFIRST {\it YJH} (green) and {\it YJHK} (red) $5{\sigma}$ photometry; and WFIRST {\it YJHK} shallower by $1.2$ mag (grey). Style conventions as in Figure \ref{fig:NIRcomp_stats}. \label{fig:wfirst}}
\end{center}
\end{figure}

The NASA WFIRST survey will have 2200 square degrees of overlap with the LSST main survey, produce significantly deeper catalogs of {\it YJH} photometry than Euclid (Table \ref{tab:maglims}), and will also have a redder (F184) filter which we will refer to as $K$. The LSST-WFIRST overlap region is projected to be ${\sim}12\%$ of the total LSST main survey of $18000$ $\rm deg^{2}$, and all galaxies detected by LSST with $i<25$ mag within $0.3<z<3.0$ would be detected by WFIRST in all four filters, {\it YJHK} (as described in Section \ref{sssec:exp_cat_demo}). WFIRST is thus expected to provide significant improvements to the photometric redshifts for the LSST galaxy catalog. For this simulation we use the same test and training set catalogs as in Section \ref{sec:euclid}, with three main differences: (1) the addition of a redder filter to the set of colors included in the Mahalanobis distance in color-space (F184, referred to here for simplicity as $K$); (2) the use of WFIRST $Y$ instead of LSST $y$ because the former is significantly deeper; and (3) the use of the WFIRST $5{\sigma}$ limiting magnitudes to simulate the NIR photometry. Cuts in apparent magnitude are applied using the $5{\sigma}$ limiting magnitudes for all filters except $i$-band, for which we use $i<25$ mag. Recall from Section \ref{sssec:exp_cat_err} that the predicted limiting magnitudes of the WFIRST image depths were recently updated to be deeper \citep{2019BAAS...51c.341D} than we have used for this simulation, so our results should be interpreted as an assessment of WFIRST's {\it minimum} impact on LSST photo-$z$ estimates.

For this simulation we do not show a true- {\it vs.} photometric-redshift plot (as in Figure \ref{fig:NIRcomp_tzpz}) because the addition of WFIRST photometry tightens up the scatter around the $z_{\rm true}=z_{\rm phot}$ locus so much -- except for the spurs at $z\lesssim0.5$ that are also seen in Figure \ref{fig:NIRcomp_tzpz} --  that the results are better demonstrated by the statistical measures. In Figure \ref{fig:wfirst} we show how the standard deviation, bias, and fraction of outliers are all drastically improved when WFIRST photometry is included. The standard deviation decreases across all redshift bins, and by ${\sim}60\%$ for redshifts $z_{\rm phot}\gtrsim1.5$, and the absolute bias is lowered to be $\lesssim0.001$ across all bins. The catastrophic outlier galaxies at low ($z_{\rm phot}<1$) and high ($z_{\rm phot}>2.5$) redshift, which are caused by degeneracies between optical color and redshift, are almost completely eliminated.

The primary reason for the improved photo-$z$ statistics is the extreme depth of the WFIRST photometry, but the additional redder filter also makes a unique contribution. In Figure \ref{fig:wfirst} we show separately the results when WFIRST filters {\it YJH} only, or {\it YJHK}, are included. We find that the additional redder filter lowers the standard deviation in the highest redshift bins ($z>2.5$) by $\sim13\%$, a small but not insignificant amount (i.e., the change is larger than the error bars in Figure \ref{fig:wfirst}). The $K$ filter has an impact at $z>2.5$ because this is where the $4000$ $\rm \AA$ Balmer break would start to influence the $H-K$ color. As a comparison, we also show the results when Euclid photometry is included. Euclid will be significantly shallower than WFIRST and so it does not improve the photo-$z$ estimates as much, but Euclid will begin operations much sooner and cover a wider area than WFIRST. Each survey will make unique contributions to cosmological analyses involving LSST photo-$z$.

The planned WFIRST survey area (${\sim}2200$ $\rm deg^2$) is significantly smaller than that of Euclid and LSST (${\sim}7200$ and ${\sim}18000$ $\rm deg^2$), but could be made wider at the expense of depth. To investigate this, we also simulate photo-$z$ estimates that include photometry from a shallow WFIRST survey for which the limiting magnitudes (Table \ref{tab:maglims}) are all reduced by 1.2 magnitudes, to $25.5$, $25.7$, $24.8$, and $24.6$ in filters {\it YJHK}, in order to extend the WFIRST survey area to cover the entire LSST footprint. The results are represented by the grey lines in Figure \ref{fig:wfirst}. We find that the standard deviation, bias, and fraction of outliers for galaxies with $z<0.5$ would be just slightly (${\sim}5$--$10\%$) lower than the results with Euclid {\it JH}. However, around $z\sim1$ the standard deviation and fraction of outliers would be nearly equivalent to the quality of a deep WFIRST survey (as is the bias for $z>1$). At high redshifts, a shallow WFIRST would reduce these statistics by at least twice as much as Euclid and, most notably, reduce the fraction of outliers to ${<}3\%$ out to $z\sim2.5$.

Given the considerable depths of the WFIRST photometry, we investigated the quality of photo-$z$ estimates based \emph{solely} on WFIRST filters {\it YJHK}. We concluded that they are not scientifically usable, and so have not shown the results in a plot. We found that the statistical measures for $z_{\rm phot}<1.5$ would be very large (standard deviation ${\sim}0.2$ and absolute bias ${\sim}0.05$), and only for $2.0<z_{\rm phot}<2.5$ does the accuracy and precision approach that of photo-$z$ from optical and NIR photometry combined (standard deviation ${\sim}0.05$ and absolute bias ${\sim}0.01$).

\subsection{A Deeper Test Set} \label{ssec:wfirst_deep}

\begin{figure}
\begin{center}
\includegraphics[width=8.5cm,trim={0cm 0cm 0cm 0cm}, clip]{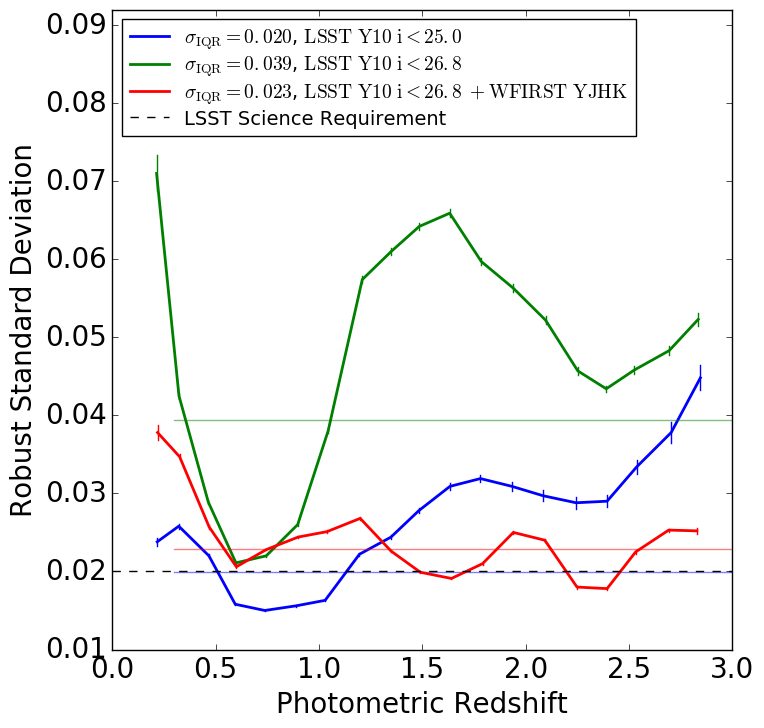}
\includegraphics[width=8.5cm,trim={0cm 0cm 0cm 0cm}, clip]{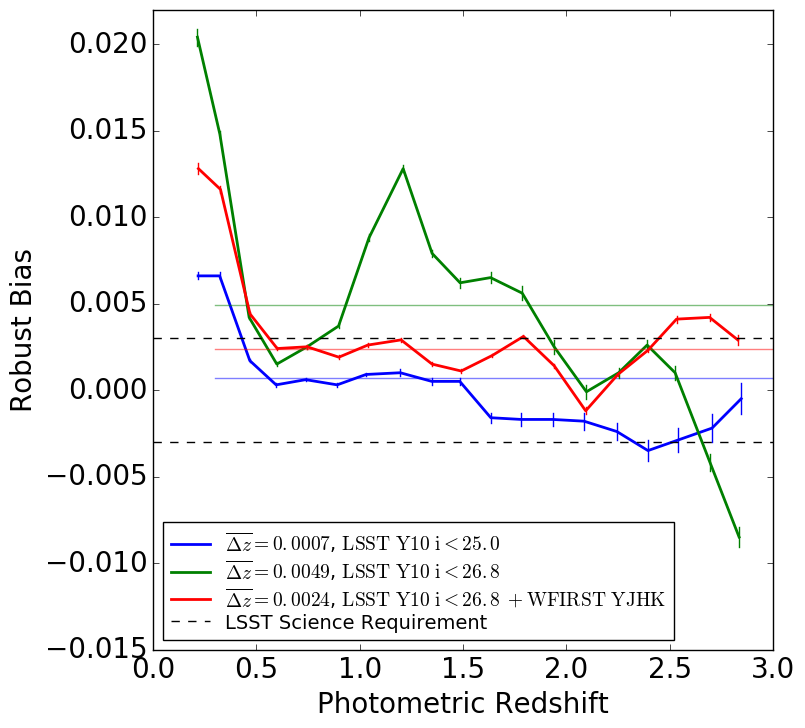}
\caption{The standard deviation (top) and bias (bottom) for photo-$z$ results for a simulated test set of galaxies with $i<25$ mag (blue), $i<26.8$ mag (green), and $i<26.8$ mag but with WFIRST {\it YJHK} photometry included (red). \label{fig:wfirstdeep}}
\end{center}
\end{figure}

The survey area of overlap between LSST and WFIRST may only be $2200$ square degrees, but this area will be observed to unprecedented depth, and that might enable alternative science goals. Here we explore the impact on photo-$z$ of adding WFIRST photometry to a deeper set of LSST galaxies that extends to $i<26.8$ mag, the LSST $i$-band $5{\sigma}$ detection limit, instead of $i<25$ mag as applied in the rest of this work. In this simulation we keep the $i<25$ mag cut applied to the training set because it is difficult to obtain large \emph{representative} samples of galaxy spectra much deeper than that. We expect that this mismatch in depth between test and training set might induce a bias in the photo-$z$ results: that the photo-$z$ are biased towards low-$z_{\rm phot}$ (under-predicted, on average) because the brighter training set has a lower average redshift. 

In all of our simulations in this work we impose the constraint that a galaxy must have three colors in order to obtain a photo-$z$ estimate, but here we furthermore constrain that those colors must be \emph{optical}. This constraint prevents the situation where the photo-$z$ results are deteriorated by a large number of faint, $i>25$ mag galaxies that are detected only (or mainly) in the WFIRST filters -- a population that would be cut from all of our other simulated test sets. This optical-colors constraint ensures we are ``comparing apples to apples" in this simulation, and evaluating only the impact of adding WFIRST photometry to an LSST galaxy catalog (and not adding WFIRST-detected galaxies to an LSST catalog).

In Figure \ref{fig:wfirstdeep} we show the standard deviation and bias for a test set of $i<26.8$ mag galaxies with LSST photometry only, and when WFIRST {\it YJHK} photometry is included. For comparison we also show the results for a test catalog with $i<25$ mag. As expected, the overall photo-$z$ results are poorer when fainter galaxies are included, but the addition of WFIRST photometry reduces the standard deviation by $\gtrsim60\%$ for $1<z_{\rm phot}<2$. As also expected, using an $i<25$ mag training set for this $i<26.8$ mag test set induced a bias in the photo-$z$ results at $z_{\rm phot}\sim1.5$ and in the highest-$z$ bins, but we find that the inclusion of the WFIRST photometry alleviates the bias. We do not show the plot for the fraction of outliers but find that, as expected, the addition of WFIRST photometry to a faint test set reduces the fraction of outliers by ${\sim}60$\% at intermediate- to high-redshifts ($z_{\rm phot} >1.5$).

In Section \ref{sec:euclid} we found that the additional degrees of freedom from adding NIR photometry could deteriorate the photo-$z$ results, especially for test sets with poorer photometric quality. Despite this simulation including fainter $i<26.8$ mag galaxies we find no such deterioration when adding WFIRST photometry, and so we do not need to apply any restrictions as was done in Section \ref{sec:euclid}.

\section{The Proposed CASTOR Mission}\label{sec:castor}

The proposed CASTOR mission would perform an imaging survey that would overlap with the LSST survey area and provide an additional passband \emph{and} deeper photometry that could be included in photo-$z$ estimates. CASTOR currently has two proposed UV/$u$ bandpass pairs, $UV$/$u_c$ and $UV_d$/$u_w$, and a $g_c$-band (as shown in the top-left panel of Figure \ref{fig:nuv}). There are several proposed surveys being considered for CASTOR, but we focus on two: (1) a primary survey that would cover the 7200 square degree region defined by the overlap of the LSST WFD and Euclid-Wide surveys (which also includes the WFIRST High Latitude Survey); and (2) the CASTOR Cadence survey that would provide multi-epoch, much deeper imaging in a ${\sim}$20 square degree region that overlaps with the LSST Deep Fields and WFIRST in the southern skies. The schedule for the CASTOR mission is not yet set, but it is not likely to launch sooner than 2027. In this analysis we only evaluate the final combined surveys' photometry. Since \citet{2018AJ....155....1G} demonstrated that LSST $u$-band data primarily helps improve the photo-$z$ quality for galaxies with $z_{\rm phot}<0.6$ and $z_{\rm phot} > 1.5$, we might expect the addition of CASTOR data to have the largest impacts at low- and high-$z$.

\subsection{Testing CASTOR Passbands for Photo-$z$}\label{ssec:castor_passbands}

\begin{figure}
\begin{center}
\includegraphics[width=8.5cm,trim={0cm 0cm 0cm 0cm}, clip]{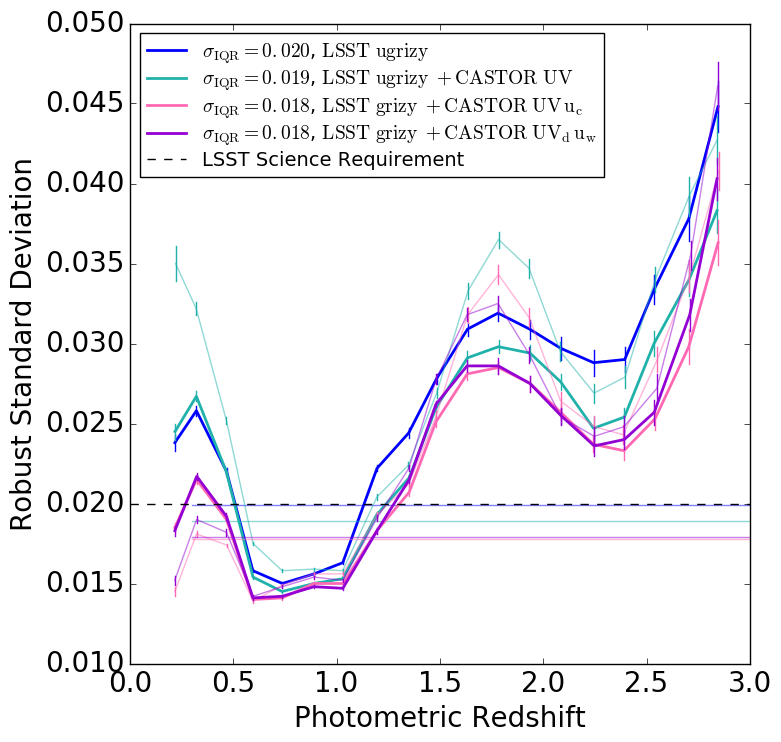}
\caption{The standard deviation as a function of redshift bin for the photo-$z$ results of using the CASTOR $UV$ passband (green), or the two proposed CASTOR bandpass pairs -- $UV$ and $u_c$ (hot pink) or $UV_d$ and $u_w$ (purple) -- compared to the LSST {\it ugrizy} filter set only (blue). Thin lines represent the results when NUV photometry is included for \emph{all} galaxies regardless of its impact on the photo-$z$ estimate. \label{fig:castor_filtsetIQR}}
\end{center}
\end{figure}

\begin{figure}
\begin{center}
\includegraphics[width=8.5cm,trim={0cm 0cm 0cm 0cm}, clip]{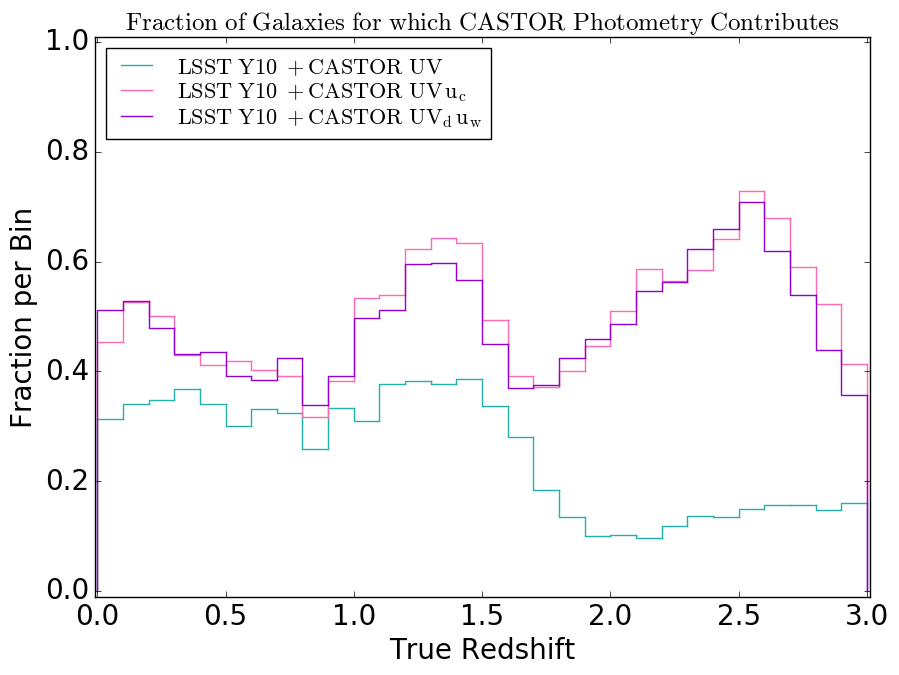}
\caption{The fraction of galaxies in each redshift bin which were brighter than CASTOR's $5{\sigma}$ limiting magnitude, i.e., $UV\leq27.4$ mag (green) and/or $u\leq27.4$ mag (pink and purple), \emph{and} for which the CASTOR photometry was included in the photo-$z$ estimate because it lowered the estimated photo-$z$ uncertainty. Shown for each of our simulations in Figure \ref{fig:castor_filtsetIQR}, as labeled in the legend. \label{fig:frac_restrict_castor}}
\end{center}
\end{figure}

To explore which $UV$/$u$ bandpass pair would be best for photometric redshifts, we simulate photo-$z$ results with the CASTOR Primary survey limits (Table \ref{tab:maglims}), imposing $i<25$ mag for the test and training set as usual. We show the standard deviation as a function of binned $z_{\rm phot}$ in Figure \ref{fig:castor_filtsetIQR}. As we found with the addition of Euclid photometry (Section \ref{sec:euclid}), including NUV photometry for \emph{all} galaxies caused a slight deterioration in the standard deviation for galaxies with $1.5<z_{\rm phot}<2$ (thin lines in Figure \ref{fig:castor_filtsetIQR}). To mitigate this we only include the CASTOR photometry if it \emph{lowers} the estimated photo-$z$ uncertainty and is thus likely to improve the photo-$z$ estimate. The statistical results after this restriction has been imposed are represented by the thicker lines in Figure \ref{fig:castor_filtsetIQR}. Figure \ref{fig:frac_restrict_castor} shows the fraction of galaxies that are detected by CASTOR and have their CASTOR photometry included in the photo-$z$ estimate. We find that the CASTOR $UV$ passband is useful for up to ${\sim}40\%$ of all simulated galaxies, and the deeper $u$-band is useful for ${\sim}40$ to $60\%$ of simulated galaxies.

The statistical measure of standard deviation shown in Figure \ref{fig:castor_filtsetIQR} demonstrates that all of the proposed CASTOR bandpasses could provide an improvement of ${\sim}10\%$ across the full redshift range. At low redshifts, $z < 0.6$, we see that the deeper $u$-band photometry provides a larger reduction in the standard deviation than the addition of the $UV$ passband alone. This is because for low-$z$ galaxies information about the Balmer break at $4000$ $\rm \AA$ is conveyed via the $u-g$ color, and the $UV-u$ color is less indicative of redshift. 

We find that, surprisingly, our restriction to only include CASTOR photometry when it decreases the photo-$z$ uncertainty actually causes a small \emph{increase} in the standard deviation in the lowest-$z$ bins, when either the CASTOR $UV$/$u_c$ or $UV_d$/$u_w$ filter pairs are used. This indicates that the CASTOR photometry can improve the reported \emph{precision} of some photo-$z$ estimates while degrading their \emph{accuracy} (this issue is also discussed in Section \ref{ssec:castor_north}). This could be mitigated by applying further conditions to our restriction, such as always including CASTOR photometry when it indicates $z_{\rm phot}<0.5$, but since the effect is minor we avoid such additional complications at this time. 

At intermediate redshifts ($1<z<1.5$) we see that all CASTOR filters provide an equivalent improvement to the standard deviation, a reduction of ${\sim}15\%$ compared to using the LSST filters alone, and that in the highest redshift bins the $UV$/$u_c$ performs a little better than $UV_d$/$u_w$. Since the $UV$/$u_c$ pair provides a slightly lower standard deviation across the full redshift range of $0.3<z<3.0$ compared to the $UV_d$/$u_w$ pair, we continue the rest of this analysis with $UV$/$u_c$.

\subsection{CASTOR Primary}\label{ssec:castor_north}

\begin{figure}
\begin{center}
\includegraphics[width=7cm,trim={0cm 0cm 0cm 0cm}, clip]{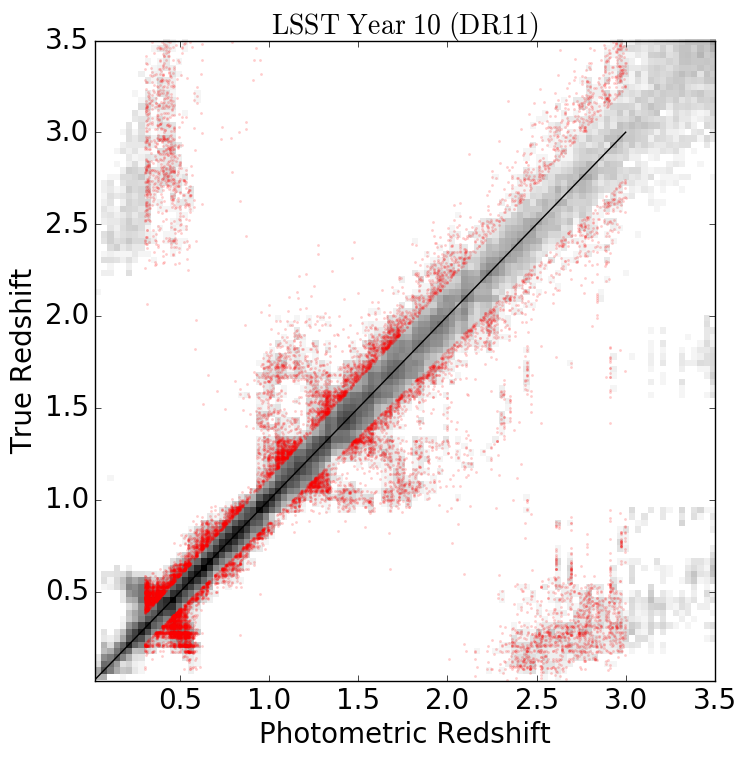}
\includegraphics[width=7cm,trim={0cm 0cm 0cm 0cm}, clip]{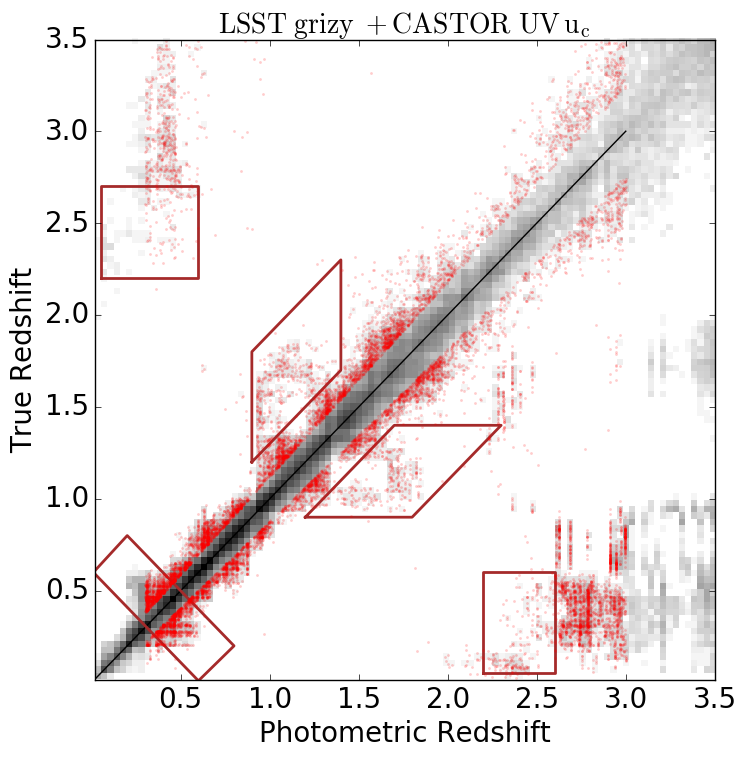}
\includegraphics[width=7cm,trim={0cm 0cm 0cm 0cm}, clip]{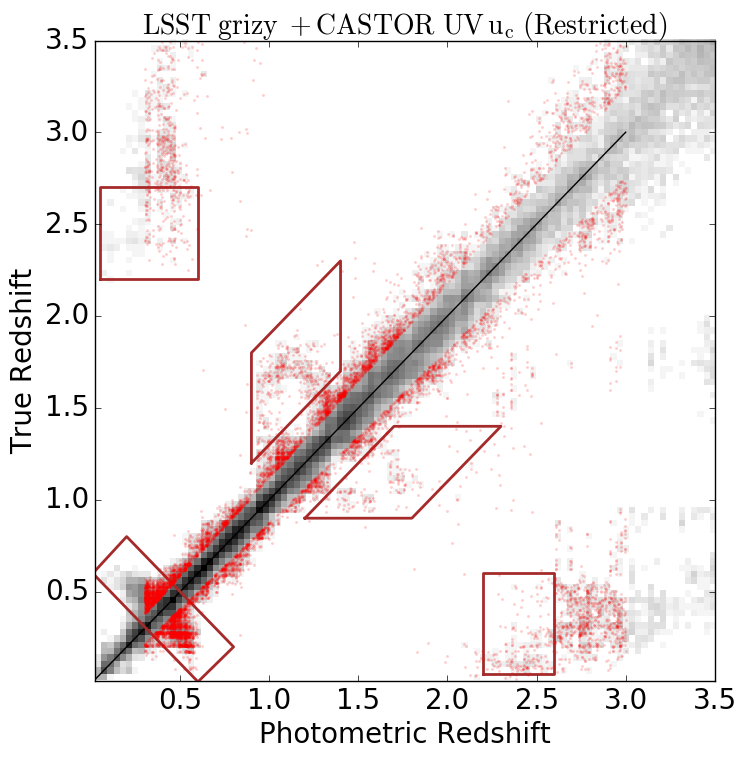}
\caption{True $vs.$ photometric redshifts for simulated galaxies with $i<25$ mag based on the LSST 10-year $5{\sigma}$ detection limits only (top); when CASTOR-Primary $UV$/$u_c$ photometry is included for all galaxies (middle); and when that CASTOR data is excluded if it increases the photo-$z$ uncertainty (bottom). Brown boxes guide the eye to sensitive outlier populations. \label{fig:castor_tzpz}}
\end{center}
\end{figure}

\begin{figure}
\begin{center}
\includegraphics[width=7.5cm,trim={0cm 0cm 0cm 0cm}, clip]{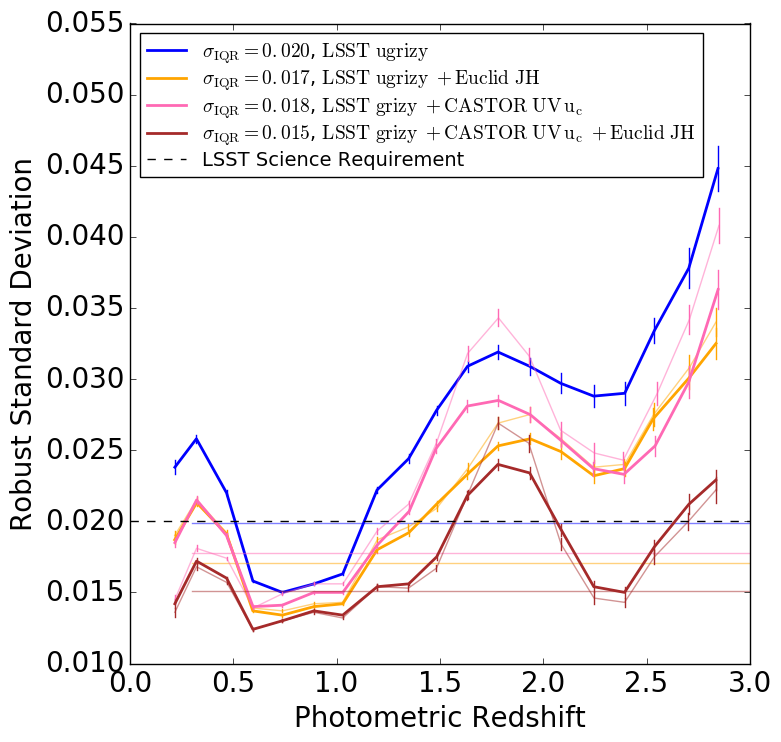}
\includegraphics[width=7.5cm,trim={0cm 0cm 0cm 0cm}, clip]{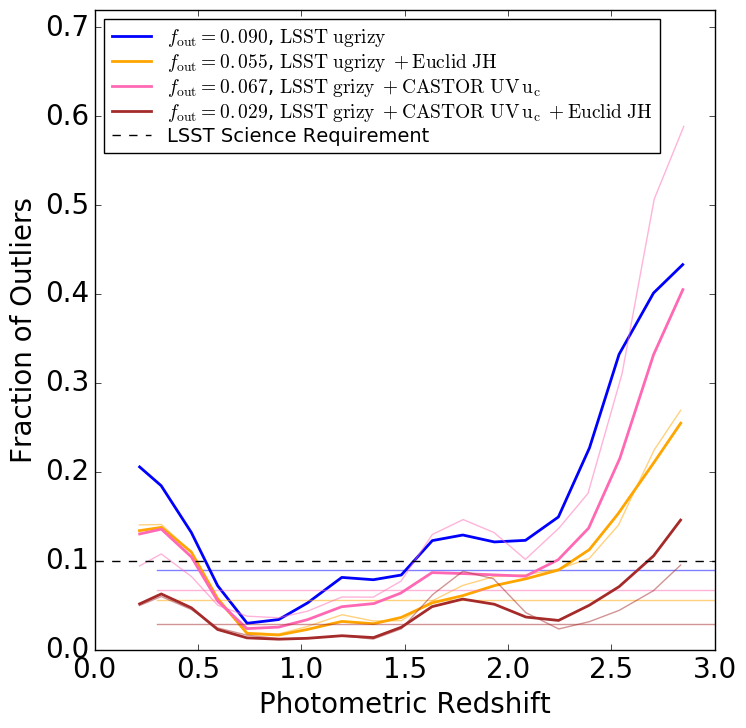}
\caption{The standard deviation (top) and fraction of outliers (bottom) of the photo-$z$ results with LSST photometry alone (blue), LSST and Euclid {\it JH} (orange), LSST and CASTOR-Primary $UV$ and $u_c$ (pink), and all three surveys combined (brown).  \label{fig:castor_stats}}
\end{center}
\end{figure}

To further investigate the potential photo-$z$ results in the CASTOR Primary survey area with the $UV$/$u_c$ passband pair, we add the CASTOR $UV$ photometry and replace the LSST $u$-band with the CASTOR $u_c$ photometry for all test- and training-set galaxies. As in the previous section, for all filters we use the surveys' $5{\sigma}$ limiting magnitudes to simulate apparent observed magnitudes and errors, and apply cuts at $i<25$ mag and the $5{\sigma}$ limits. We do not replace the LSST $g$-band with the CASTOR $g_c$-band because, as shown in Table \ref{tab:maglims}, the 10-year LSST main survey is predicted to be slightly deeper than the CASTOR Primary survey in $g$-band.

In Figure \ref{fig:castor_tzpz} we show three panels representing the true {\it vs.} photometric redshifts when: (1) only LSST {\it ugrizy} photometry is used; (2) the CASTOR-Primary $UV$ and $u_c$ photometry is included for all galaxies; and (3) the CASTOR photometry is included with the restriction that it must decrease the photo-$z$ uncertainty (as described in Section \ref{ssec:castor_passbands}). By comparing the bottom two panels with the top panel, we can immediately see how adding CASTOR data resolves degeneracies in galaxy color-space that cause outliers in certain regions of the $z_{\rm true}$--$z_{\rm phot}$ plane (brown boxes). By comparing just the bottom two panels we can see that our restriction inhibits the reduction of outliers in the low-$z$ spurs at $z<0.5$, indicating that in some cases the CASTOR photometry improves the photo-$z$ accuracy but not its precision, as discussed above.

The statistical measures of photo-$z$ quality are shown in Figure \ref{fig:castor_stats}, which plots the standard deviation and fraction of outliers in bins of photo-$z$. The bias is not significantly impacted by CASTOR photometry and so is not shown. As described in the previous section, thin lines represent results when the CASTOR photometry is included for \emph{all} galaxies, without the restriction that the addition must result in a smaller photo-$z$ uncertainty. 

The top panel of Figure \ref{fig:castor_stats} illustrates how adding CASTOR to LSST photometry reduces the standard deviation by ${\sim}10\%$ for galaxies with $0.3<z_{\rm phot}<3.0$. At low redshifts ($z_{\rm phot}\lesssim0.6$) we find that the impact of CASTOR is equivalent to that of Euclid, but that applying CASTOR photometry without restriction is more impactful and could reduce the standard deviation by ${\sim}30\%$. At intermediate redshifts, $1<z_{\rm phot}<2$, we see that CASTOR provides a reduction of ${\sim}15\%$ in standard deviation compared to using the LSST filters alone. In the redshift range $1<z<1.5$ the $UV$ filter is passing over the UV-upturn from evolved hot stars \citep[e.g.,][]{1990ApJ...364...35G,1999MmSAI..70..691G}, and the flux from this UV-upturn provides redshift information via the $UV-u$ color. By redshift $z\sim1.5$ the Lyman-$\alpha$ break has entered the $UV$ bandpass, but the $UV-u$ color continues to provide redshift information until $z\sim2$. At intermediate redshifts ($1.4<z_{\rm phot}<2$) we find that the addition of Euclid is more effective at reducing the standard deviation.

In even the highest redshift bins we continue to see an improvement to the standard deviation when CASTOR $UV$ photometry is included, despite the fact that Figure \ref{fig:frac_restrict_castor} shows that the $UV$ only contributes to the photo-$z$ estimate ${\sim}10\%$ of the time for galaxies with \emph{true} redshifts $z_{\rm true}>2$. This is the result of the CASTOR photometry assisting with the identification of truly lower-$z$ galaxies that have optical colors which are degenerate with higher-$z$ galaxies. These galaxies would otherwise be assigned a high photo-$z$ based on their optical colors alone, where they increase the scatter in the high-$z_{\rm phot}$ bins. In the bottom panel of Figure \ref{fig:castor_stats} we show that the improvements to the fraction of outliers are generally of a similar scale, except that CASTOR could reduce this metric by ${\sim}50\%$ for $z<0.5$, and Euclid is more impactful for all galaxies with $z>1$. 

Generally, for both statistical measures and across all redshift bins, Figure \ref{fig:castor_stats} demonstrates how the addition of \emph{both} CASTOR and Euclid could provide further improvements than either could on its own. This indicates that the correlations between galaxy UV and NIR photometry and redshift, which are put to use when we add CASTOR and Euclid data to the photo-$z$ estimates, are independent of each other. The fact that each survey provides a similar impact on the photo-$z$ results when added individually is just a coincidence, and CASTOR and Euclid would each deliver complementary information to the photo-$z$ estimates.

\subsection{CASTOR Cadence}\label{ssec:castor_south}

The proposed CASTOR Cadence survey would cover $20$ square degrees and might overlap with the LSST deep drilling fields, Euclid's deep fields, and/or the WFIRST survey area. Similar to Section \ref{ssec:euclid_deeptrain}, we evaluate the impact of using this region to build a deeper spectroscopic training set. For this simulation we use a test set of galaxies limited to $i<25$ mag, with photometric quality based on the $5{\sigma}$ limiting magnitudes of a 10-year LSST survey, the CASTOR Primary survey, and/or the Euclid main survey (all as quoted in Table \ref{tab:maglims}). We use a training set of galaxies also limited to $i<25$ mag, with photometric quality based on the $5{\sigma}$ limiting magnitudes of a LSST deep drilling field (Section \ref{ssec:euclid_deeptrain}), the CASTOR Cadence deep field ($m_{\rm UV} \sim 29.25$, $m_u \sim 28.95$, and $m_g \sim 28.45$ mag; Section \ref{sec:intro}), and the WFIRST survey (Table \ref{tab:maglims}). Two aspects that are different about this simulation from all preceding it are that (1) the CASTOR $g_c$ is replacing LSST $g$-band for the first time because the Cadence survey would be deeper than a stacked LSST deep-drilling field, and (2) the two NIR surveys' photometry is being mixed, with WFIRST used for the training set and Euclid for the test set.

\begin{figure}
\begin{center}
\includegraphics[width=7.5cm,trim={0cm 0cm 0cm 0cm}, clip]{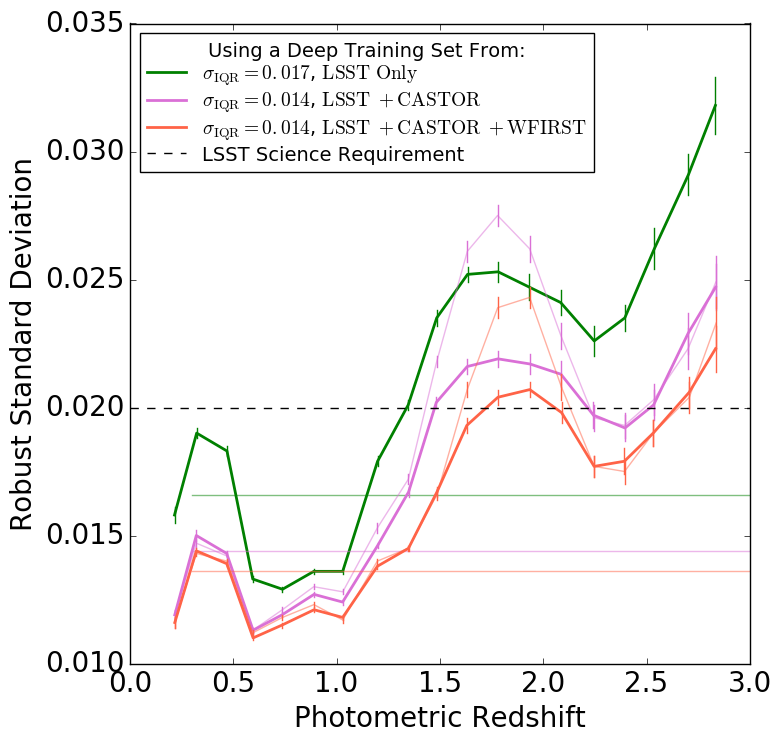}
\includegraphics[width=7.5cm,trim={0cm 0cm 0cm 0cm}, clip]{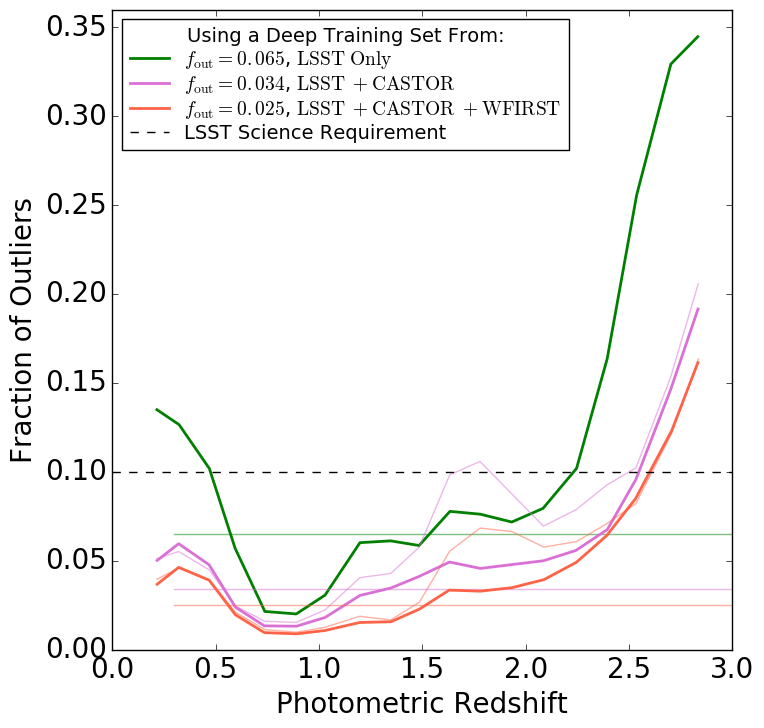}
\caption{The standard deviation (top) and fraction of outliers (bottom) of the photo-$z$ results when a deep training set that includes photometry from LSST (green), CASTOR (pink), or both CASTOR and WFIRST (red) is used. \label{fig:castorDT_stats}}
\end{center}
\end{figure}

In Figure \ref{fig:castorDT_stats} we demonstrate the impact of using a deeper training set based on the LSST deep drilling fields only, when the CASTOR Cadence photometry is included, and when the WFIRST photometry is also included. We can see that a deeper training set that includes CASTOR photometry can significantly decrease the standard deviation and fraction of outliers at low ($z_{\rm phot}<1.5$) and high ($z_{\rm phot}>2$) redshifts. As expected, including the WFIRST photometry in this deep training set provides further improvements across all redshifts.

\section{Discussion and Conclusions}\label{sec:conc}

In this study we have investigated the impact of including photometry from the Euclid, WFIRST, and/or CASTOR space telescopes surveys' in the photometric redshift estimates for LSST galaxies. Using mock galaxy catalogs with simulated photometry we have measured the robust standard deviation, robust bias, and fraction of outliers for our results. In Table \ref{tab:sum} we summarize the \emph{change} in these statistics for three main redshift ranges: low ($0.3<z_{\rm phot}<0.6$), intermediate ($0.8<z_{\rm phot}<1.2$), and high ($2.0<z_{\rm phot}<2.5$). The table's values are the ratio of the statistical results with NIR or NUV data to the results when the LSST 10-year photometry alone is used to estimate photometric redshifts. For example, the factor representing the decrease in the robust standard deviation within the redshift range $0.3<z_{\rm phot}<0.6$ when Euclid photometry is included is $\sigma_{\rm LSST+Euclid}/\sigma_{\rm LSST} = 0.87\pm0.01$. The values quoted in Table \ref{tab:sum} are based on simulations in which the inclusion of NIR and NUV photometry for Euclid and CASTOR is subject to the restrictions discussed in Sections \ref{sec:euclid} and \ref{sec:castor}.

\begin{table} 
\begin{center} 
\caption{Summary of Statistical Results} 
\label{tab:sum} 
\begin{tabular}{|c|ccc|} 
\hline 
\hline 
 & \multicolumn{3}{c|}{\it Fractional change$^{a}$ from using LSST$^{b}$} \\ 
$z_{\rm phot}$ & \multicolumn{3}{c|}{\it \smash{\raise2pt\hbox{photometry only when we add:}} } \\
Range & Euclid$^{c}$ & WFIRST & CASTOR$^{d}$ \\
\hline 
 & \multicolumn{3}{l|}{\it Robust Standard Deviation:} \\ 
0.3--0.6 &  0.87 $\pm$ 0.01  &  0.74 $\pm$ 0.01  &  0.86 $\pm$ 0.01  \\  
0.8--1.2 &  0.87 $\pm$ 0.01  &  0.75 $\pm$ 0.01  &  0.93 $\pm$ 0.01  \\  
2.0--2.5 &  0.81 $\pm$ 0.02  &  0.32 $\pm$ 0.01  &  0.83 $\pm$ 0.02  \\  
\hline 
 & \multicolumn{3}{l|}{\it Robust Bias:} \\ 
0.3--0.6 &  1.45 $\pm$ 0.12  &  0.88 $\pm$ 0.08  &  0.89 $\pm$ 0.09  \\  
0.8--1.2 &  0.97 $\pm$ 0.31  &  0.63 $\pm$ 0.23  &  0.30 $\pm$ 0.22  \\  
2.0--2.5 &  0.29 $\pm$ 0.12  &  0.09 $\pm$ 0.05  &  0.73 $\pm$ 0.16  \\  
\hline 
 & \multicolumn{3}{l|}{\it Fraction of Outliers:} \\ 
0.3--0.6 &  0.83  &  0.47  &  0.79  \\  
0.8--1.2 &  0.46  &  0.15  &  0.68  \\  
2.0--2.5 &  0.57  &  0.01  &  0.62  \\  
\hline 
\multicolumn{4}{l}{(a) E.g., $\sigma_{\rm LSST+Euclid} / \sigma_{\rm LSST}$.} \\
\multicolumn{4}{l}{(b) A 10-year LSST survey.} \\
\multicolumn{4}{l}{(c) Using the Euclid $1{\sigma}$ detection limit.} \\
\multicolumn{4}{l}{(d) Using CASTOR-Primary $UV$ and $u_c$ filters.}
\end{tabular} 
\end{center} 
\end{table} 

\begin{table} 
\begin{center} 
\caption{Summary of Results by Sky Area} 
\label{tab:sumzones} 
\begin{tabular}{|c|ccc|} 
\hline 
\hline 
 & \multicolumn{3}{c|}{\it Fractional change$^{a}$ from results in} \\ 
$z_{\rm phot}$ & \multicolumn{3}{c|}{\it \smash{\raise2pt\hbox{LSST Zone 0$^{b}$ for sky area in:}} } \\
Range  & Zone 1$^{c}$ & Zone 2$^{d}$ & Zone 3$^{e}$ \\ 
\hline 
 & \multicolumn{3}{l|}{\it Robust Standard Deviation:} \\ 
0.3--0.6 &  0.72 $\pm$ 0.01  &  0.63 $\pm$ 0.01  &  1.57 $\pm$ 0.02  \\  
0.8--1.2 &  0.83 $\pm$ 0.01  &  0.74 $\pm$ 0.01  &  2.12 $\pm$ 0.02  \\  
2.0--2.5 &  0.57 $\pm$ 0.01  &  0.62 $\pm$ 0.02  &  1.82 $\pm$ 0.05  \\  
\hline 
 & \multicolumn{3}{l|}{\it Robust Bias:} \\ 
0.3--0.6 &  0.99 $\pm$ 0.09  &  0.02 $\pm$ 0.04  &  0.99 $\pm$ 0.14  \\  
0.8--1.2 &  0.33 $\pm$ 0.21  &  0.92 $\pm$ 0.27  &  1.23 $\pm$ 0.57  \\  
2.0--2.5 &  0.56 $\pm$ 0.11  &  0.27 $\pm$ 0.10  &  3.01 $\pm$ 0.53  \\  
\hline 
 & \multicolumn{3}{l|}{\it Fraction of Outliers:} \\ 
0.3--0.6 &  0.36  &  0.30  &  0.90  \\  
0.8--1.2 &  0.29  &  0.22  &  2.00  \\  
2.0--2.5 &  0.23  &  0.31  &  1.22  \\  
\hline 
\end{tabular}
\begin{tabular}{l}
(a) E.g., $\sigma_{\rm Zone\ 1} / \sigma_{\rm Zone\ 0}$. \\
(b) Zone 0: LSST only (${\sim}$18000 deg${^2}$). \\
(c) Zone 1: LSST, Euclid, and CASTOR (${\sim}$7200 deg${^2}$). \\
(d) Zone 2: LSST, WFIRST, and CASTOR (${\sim}2000$ deg${^2}$). \\
(e) Zone 3: Shallow LSST, Euclid, and CFHT-$u$ (${\sim}3000$ deg${^2}$).
\end{tabular} 
\end{center} 
\end{table} 

For almost every statistic and every redshift range, we find a significant improvement to the photo-$z$ results when the Euclid, WFIRST, or CASTOR photometry is included, and the improvements tend to be greater in the highest-redshift bins (i.e., the factors are smaller). In the one and only case where this factor is greater than $1$, the robust bias when Euclid is added in the lowest-$z$ bin has a very low absolute value, $0.0025$, which is well within the LSST SRD's targeted range (see Figure \ref{fig:NIRcomp_stats}). 

In this study we have also explored the relative photo-$z$ results in four main sky-area zones of overlap between LSST, Euclid, WFIRST, and CASTOR: (0) the ${\sim}18000$ square degrees of the LSST wide-fast-deep survey at 10 years with filters {\it ugrizy}; (1) the ${\sim}7200$ square degrees of overlap between LSST {\it grizy}, Euclid-Wide {\it JH}, and CASTOR-Primary $UV$ $u_c$; (2) the ${\sim}2000$ square degrees of overlap between LSST {\it griz}, WFIRST {\it YJHK}, and CASTOR-Primary $UV$ $u_c$; and (3) the northern area of ${\sim}3000$ square degrees with shallow coverage from LSST {\it griz}, Euclid {\it YJH}, and CFHT-$u$. In Table \ref{tab:sumzones} we summarize the fractional changes in each statistical measure, in three redshift ranges, for galaxies in sky-area zones 1, 2, and 3 compared to zone 0 (i.e., areas of overlap with Euclid, WFIRST, and/or CASTOR compared to areas with LSST alone). 

We find that the overall improvements to the standard deviation and fraction of outliers in zones 1 and 2 are quite similar, while the improvements to the bias appear to be quite different in the low- and intermediate-redshift ranges for zones 1 and 2. For example, it seems that the bias at low-$z$ is barely improved in zone 1 but improved by almost two orders of magnitude in zone 2. However, the bias in photo-$z$ results for zones 0, 1, and 2 is low, within the SRD's targeted ranges, and so the fractional improvements in Table \ref{tab:sumzones} represent small changes to already-small values.  As expected, the photo-$z$ results in zone 3 -- the potential shallow northern LSST survey extension discussed in Section \ref{ssec:euclid_north} -- are mostly of a poorer quality compared to zone 0. The exception is in the lowest-$z$ bin, where the the standard deviation and fraction of outliers are not deteriorated, because the addition of Euclid and CFHT photometry has mitigated the impact of shallower LSST photometry.

Below we summarize the results for each survey in turn, and then discuss some ideas for future work regarding ways in which space-based imaging could further improve to the LSST photo-$z$ results.

\subsection{Euclid}\label{ssec:conc_euclid}

We have simulated photometric redshift results for LSST galaxies that overlap with the Euclid-Wide survey area. We found that Euclid mainly improves the photo-$z$ estimates at redshifts $z>1$, as expected, because this is where the Balmer break is redshifted beyond the optical filters. Quantitatively, we found that the addition of Euclid $5{\sigma}$ detections to the LSST 10 year catalog can reduce the standard deviation for galaxies with $z>1$ (or within the full redshift range of $0.3<z<3.0$) by ${\sim}20\%$ (${\sim}10\%$), and the fraction of outliers by ${\sim}40\%$ (${\sim}25\%$). We showed that Euclid would have a relatively larger positive impact on the photo-$z$ estimates when added to the 10-year LSST photometry than at earlier years of the survey, and/or when Euclid detections down to $1{\sigma}$ are included (instead of $5{\sigma}$). We conclude that reducing the standard deviation in the photo-$z$ results is more efficiently done with deeper LSST {\it ugrizy} photometry than with Euclid data, but emphasize that Euclid does offer the unique benefits of reducing the standard deviation at $z>1$ and especially of reducing the fraction of outliers -- and furthermore point out that observing longer with LSST $z$ and $y$ filters would not recreate the benefits of including Euclid photometry. We demonstrated how a spectroscopic training set from overlapping LSST and Euclid deep drilling fields can significantly improve the photo-$z$ results, and how the photo-$z$ for a shallow northern LSST extension that overlaps with the Euclid-Wide survey would be scientifically useful (i.e., would have an average standard deviation just ${\sim}2$ times higher than the LSST year 10 survey).

\subsection{WFIRST}\label{ssec:conc_wfirst}

Compared to Euclid, WFIRST will provide deeper NIR photometric catalogs and one additional, redder filter (F184, which we have referred to as $K$), but will have a significantly smaller overlap area with LSST. Within the overlapping area we have demonstrated that the addition of WFIRST NIR photometry would provide drastically improved photo-$z$: the standard deviation is reduced by $\gtrsim50\%$ (${\sim}25\%$) for galaxies with $z>1.5$ ($0.3<z<3$), the bias becomes negligible for $z>1$ (and is reduced by ${\sim}30\%$ for $z<0.5$), and the fraction of outliers is reduced to just ${\sim}2\%$ for galaxies within $0.3<z<3.0$. In particular, the catastrophic outliers caused by a degeneracy between optical colors of galaxies at low- and high-redshifts are almost entirely removed. We also considered the impact of adding WFIRST to a deeper LSST catalog ($i<26.8$ instead of $i<25$ mag), and found that WFIRST would provide critical improvements to the standard deviation, lowering it by $50\%$ for intermediate redshifts.

\subsection{CASTOR}\label{ssec:conc_castor}

The proposed CASTOR mission might provide deep $UV$-, $u$-, and $g$-band photometry that overlaps with the LSST main survey. We found that including the CASTOR-Primary survey photometry improves the LSST photo-$z$ estimates at all redshifts, as expected, because detections in the bluer passbands help to break degeneracies in the optical colors of low and high redshift galaxies by identifying galaxies that are truly low-$z$. In particular, we showed that including CASTOR photometry could reduce the standard deviation by ${\sim}30\%$ and the fraction of outliers by ${\sim}50\%$ at $z<0.5$. We also demonstrate how further improvements to the photo-$z$ estimates might be attained by building a deeper spectroscopic training set from a combination of the CASTOR-Cadence and LSST deep drilling fields. Although we illustrated how both of CASTOR's potential $UV$- and $u$-band pairs could provide similar benefits to the LSST photo-$z$, at the time of this publication the CASTOR passbands and surveys were still in the proposal stage and might continue to evolve from what we have assumed in this work.

\subsection{Summary}\label{ssec:conc_sum}

Generally, we have found that increased photometric depth provides the largest potential reduction in the standard deviation for predicted LSST photo-$z$ estimates, compared to including additional filters beyond the optical. However, we have also found that the addition of UV and/or NIR filters can significantly reduce the standard deviation in particular redshift ranges and, more importantly, that additional filters are absolutely necessary for reducing the fraction of $3{\sigma}$ and catastrophic outliers. We have demonstrated how the addition of NIR or UV photometry from Euclid or CASTOR, respectively, would each result in similar reductions for the standard deviation and fraction of outliers when included in LSST photo-$z$ estimates individually. In addition, we have shown how Euclid and CASTOR each provide \emph{complementary} information about a galaxy's redshift, and thus how additional significant improvement to the photo-$z$ estimates could be obtained by including data from both surveys. As a final note, the deeper NIR photometry from WFIRST should deliver superb photo-$z$ estimates for LSST galaxies, albeit in the smaller area of overlap between the LSST and WFIRST surveys.

\subsection{Future Work}\label{ssec:conc_future}

The Euclid, WFIRST, and CASTOR missions are all space telescopes, and there are additional merits to their data aside from just the improved photometric quality and expanded spectral range (e.g., spatial resolution). These aspects can also be used to improve the LSST photo-$z$, over and above what we have demonstrated (e.g., \citealt{2019A&A...621A..26P}). Doing so requires a photo-$z$ estimator that has been constructed to ingest, and properly apply, the additional features of the data. Here we suggest several options for obtaining even greater improvements for LSST photo-$z$ from the Euclid, WIFRST, and/or CASTOR surveys, which are beyond the scope of this work.

\smallskip
\begin{itemize}
\item With the CMNN estimator, all galaxies must have a photometric detection with appropriate errors in order to simulate photometric redshifts. A photo-$z$ estimator which fully utilized non-detections might be better able to quantify the improvements offered by surveys that are significantly shallower than the LSST.
\item With the CMNN estimator, adding extra filters can deteriorate the photo-$z$ quality if the photometry is not constraining enough to balance the additional degrees of freedom. In this work we have mitigated this issue by only adding extra filters when they're likely to improve the photo-$z$ estimate. A photo-$z$ estimator that takes a more sophisticated approach to the provisional addition of photometry might be able to show even greater improvement in the photo-$z$ quality.
\item Photometric differences between filters that are similar, such as LSST's $y$ and the Euclid or WFIRST $Y$, could be used to constrain the redshift of strong emission lines and thus provide more precise photo-$z$ for a subset of galaxy types.
\item Euclid and WFIRST will have grism data, which we have not considered in this work, but which could be used to further improve the photo-$z$ estimates -- either by providing additional SED information for test galaxies, or perhaps enhancing the training set photometry.
\item All space-based imaging surveys offer significantly better spatial resolution than ground-based imaging, and this could improve the photo-$z$ by, for example, providing galaxy size and shape priors and/or improving the treatment of blended objects. This may require pixel-level analysis, whereas here we have worked only with mock catalogs.
\end{itemize}

\section*{Acknowledgements}

This material is based upon work supported in part by the National Science Foundation through Cooperative Agreement 1258333 managed by the Association of Universities for Research in Astronomy (AURA), and the Department of Energy under Contract No. DE-AC02-76SF00515 with the SLAC National Accelerator Laboratory. Additional funding for Rubin Observatory comes from private donations, grants to universities, and in-kind support from LSSTC Institutional Members.
AJC acknowledges support by the U.S. Department of Energy, Office of Science, under Award Number DE-SC-0011635.
SJS acknowledges support from DOE grant DE-SC0009999 and NSF/AURA grant N56981C.
MLG, AJC, CBM, ZI, SFD, RLJ, MJ, PY, and JBK acknowledge support from the DiRAC Institute in the Department of Astronomy at the University of Washington. The DiRAC Institute is supported through generous gifts from the Charles and Lisa Simonyi Fund for Arts and Sciences, and the Washington Research Foundation.
MJ wishes to acknowledge the support of the Washington Research Foundation Data Science Term Chair fund, and the University of Washington Provost?s Initiative in Data-Intensive Discovery.

\bibliography{ms}

\end{document}